\documentstyle[12pt]{article}
\def\lromn#1{\uppercase\expandafter{\romannumeral#1}}

\begin{document}

\begin{flushright}
TU/97/518 \\
\end{flushright}

\vspace{12pt}

\begin{center}
\begin{Large}

\renewcommand{\thefootnote}{\fnsymbol{footnote}}
\bf{
Elementary Processes in Dissipative Cosmic Medium }
\footnote[2]
{
Lecture given at Workshop on The Cosmological Constant
and Evolution of the Universe, held at Tokyo, December 25-27, 1996.
To appear in the Proceedings.
}

\end{Large}

\vspace{36pt}
\begin{large}
M. Yoshimura

Department of Physics, Tohoku University\\
Sendai 980-77 Japan\\
\end{large}

\end{center}

\begin{center}
\vspace{54pt}

{\bf ABSTRACT}
\end{center}

Quantum dynamics of a finite degrees of freedom is often much
affected by interaction with the larger environment of cosmic medium.
In this lecture we first review some recent developments of the
theory of quantum dissipation in the linear open system.
In the second part we discuss two classes of applications:
decay of unstable particle in medium, and environmental effect
on the parametric particle production.
The first subject of particle decay may have important consequences
on the scenario of baryogenesis due to a GUT heavy particle.
On the other hand, the parametric resonant particle production
is related to the reheating problem after inflation, and its
environmental effect is related also,
perhaps more importantly, to the decay of moduli fields
in supergravity models.


\newpage

\begin{center}
\begin{Large}
{\bf \lromn 1 Introduction}
\end{Large}
\end{center}

Ideal elementary process rarely occurs, 
presumably except in carefully prepared laboratory experiments.
It might be useful to recall that even processes usually considered 
elementary can be regarded  taking place in complex environments.
This has to do with how one separates a system in question from a
surrounding environment.
For instance, the beta decay, the fundamental weak process of
\( \:
n \:\rightarrow  \: p + e + \bar{\nu} _{e}
\: \),
when it occurs in nuclei, is compounded by effects of nuclear strong 
interaction with the rest of nucleons.
Presumably the most spectacular of this kind is how nucleon decay proceeds.
The extremely weak process of baryon number violating process at the
quark-lepton level must inevitably occur in the hadronic environment.
A repeated question on the nucleon decay raised in the past is 
how the extremely slow process
of the lifetime of order $10^{31}$ years or even larger is modified
by strong interaction having the time scale of order, $10^{-23}$ seconds.
Thus one must deal with the short time limit of the decay law, and
one must face the fact that the exponential decay law is not
exact in quantum mechanics.
To my knowledge even now there is no convincing calculation 
of the nucleon decay rate, 
assuming that the law of grand unified theory is known.

In cosmology essentially every elementary process occurs in macroscopic
cosmic medium.
The process needs not involve a finite number of fundamental particles alone.
For instance, one can consider a macroscopic quantum process 
such as the first order phase transition taking place in thermal
environment.
Here the order parameter of the symmetry, the magnitude of a Higgs field,
is the system variable and it couples to essentially all light particles
of the standard model that make up thermal environment.
With these in mind, in cosmology there are plenty of places one has 
to estimate effects of medium. 
When one has a clear idea of how to separate the "small" system 
from the environment,
the essential part of the problem becomes how elementary processes are
modified in dissipative medium.
We feel that the problem should be analyzed from the first
principles of quantum mechanics, at least at the conceptual level.
In condensed matter physics this problem is generically known as
quantum Brownian motion or quantum dissipation.

It has become increasingly clear to us that the bulk of the past works
on quantum dissipation relies on the simplified form of quantum friction,
namely dissipation which is local in time.
This corresponds to the exponential decay law when one examines
the fate of some initial excitation.
This approximation is excellent in the most dominant phase of
dissipation.
But it fails both in early time and very late time behavior of
the quantum system immersed in dissipative medium.
Some formalism exists to deal with more general nonlocal dissipation, but
they are not very useful to the problems we would like to address.
Indeed, some of our new results have not been recognized in the past.

In this talk I shall first review what we have recently achieved 
with regard to the general framework of quantum dissipation.
Then in the second part I shall discuss two applications of this formalism:
particle decay process in thermal medium and parametric resonant effect
in medium.
These are related to some important problems in modern cosmology;
baryogenesis, inflation, and the moduli problem in supergravity models.

Most results described here are already in our papers,
\cite{hjmy 96}, \cite{jmy-96-1}, \cite{jmy-96-2}.
But I reorganize arguments with a new perspective, and
add some new material.
I attempted in this written version to explain basic results in a
self-contained manner.

\vspace{1cm} 
\begin{center}
\begin{Large}
{\bf \lromn 2 Basic Formalism of Quantum Dissipation}
\end{Large}
\end{center}

There are two standard methods for the theory of quantum dissipation.
The first one is the approach based on quantum Langevin equation
\cite{q-dissipation langevin}.
The other is the path integral approach initiated by Feynman and Vernon
\cite{feynman-vernon}, \cite{q-dissipation path}, \cite{qbm path review},
\cite{jmy-96-1}.
Both methods have merits and demerits, and one should use both
in clarifying specific details of relevant problems.
Before going to explain some fundamentals of these approaches,
let me explain how one models the dissipation for a small quantum system,
being caused by interaction with a larger environment.

The basic idea taken by most approaches is that existence of
a continuously many environment variables coupled to a finite number of 
subsystem variables is the essential part of dissipation.
After all one does not know all details of the environment.
Indeed, ignorance should be the privilege here.
It is fortunate that dissipation, or at least something recognized as
such, occurs once one makes no measurement of the environment.
Thus one postulates that detailed modeling of the environment and
its coupling to the small system should be unimportant to
dissipative behavior of the system.
After integrating out the environment variables, one should have
only a few phenomenological parameters to describe dissipation,
ultimately obtainable from experiments.
Despite of this phenomenological nature one should base all auguments
on rigorous quantum physics, and the modeling is inevitable.

One is thus led to the theory of the simplest, yst the most fundamental
model of quantum dissipation, the linear dissipation in open systems.
In this approach one models the environment by a continuously many set of
harmonic oscillators of some arbitrary spectrum and couples it to
the subsystem via a bilinear term.
Let the subsystem variable in question be denoted by $q$ and 
the environment variable by $Q_{a}$. 
For simplicity we assume that the subsystem has one degree of freedom,
but it should be evident to extend it to any finite degrees of freedom.
We often use a discrete label for the environment variable, but
actually we assume that there are continuously many of them.
The Lagrangian of our problem consists of three parts: 
\begin{equation}
L = L_{q}[q] + L_{Q}[Q] + L_{{\rm int}}[q\,, Q] \,.
\end{equation}
We take for the system-environment interaction the bilinear term:
\begin{eqnarray}
&&
L_{q} = \frac{1}{2}\, \dot{q}^{2} - V(q) \,, 
\\ &&
L_{Q} = \frac{1}{2}\, \sum_{a}\,(\,\dot{Q}_{a}^{2} - \omega _{a}^{2}\,
Q_{a}^{2}\,) \rightarrow 
\frac{1}{2}\, \int_{\omega _{c}}^{\infty }\,d\omega \,\left( \,
\dot{Q}^{2}(\omega ) - \omega ^{2}\,Q^{2}(\omega ) \,\right)
\,,
\\ &&
L_{{\rm int}} = -\, q\,\sum_{a}\,c_{a}Q_{a} \rightarrow 
-\,q\,\int_{\omega _{c}}^{\infty }\,d\omega \,c(\omega )Q(\omega )
\,.
\end{eqnarray}
$\omega _{c}$ taken to be positive is the smallest of the environment
frequency spectrum.

As a technical aside, the renormalization of the frequency will be 
discussed later in appropriate places.
Introduction of a counter term in relation to the renormalization
as often discussed in the literature is actually the problem of how
to relate the parameters of the theory to observable quantities. 

The strategy of subsequent investigation is to integrate out quantum
dynamics of the environment, by averaging over its initial ensemble
given by some mixed state.

\vspace{1cm} 
{\bf \lromn 2A Operator method}

\vspace{0.5cm} 
The Heisenberg equation of motion for the subsystem and the environment
variable is derived from the above Lagrangian,
\begin{eqnarray}
&&
\stackrel{..}{q} + V'(q) = -\,\int_{\omega _{c}}^{\infty }\,d\omega \,
c(\omega )Q(\omega \,, t) \,, 
\\ &&
\stackrel{..}{Q}(\omega \,, t) + \omega ^{2}\,Q(\omega \,, t) 
= -\,c(\omega )q
\,.
\end{eqnarray}
The environment part is first solved as a sum of the homogeneous and
the inhomogeneous solutions:
\begin{eqnarray}
&&
Q(\omega \,, t) = Q_{i}(\omega )\,\cos (\omega t) + P_{i}(\omega )\,
\frac{\sin (\omega t)}{\omega } \nonumber \\ &&
\hspace*{0.5cm} 
-\,\frac{c(\omega )}{2\omega }\,\int_{0}^{t}\,d\tau \,\sin \omega (t - \tau )
\,q(\tau ) - \frac{c(\omega )}{2\omega }\,\int_{t}^{t_{f}}\,d\tau \,
\sin \omega (\tau - t)\,q(\tau ) 
\nonumber \\ &&
\hspace*{0.5cm} -\,
\frac{c(\omega )}{2\omega }\,\int_{0}^{t_{f}}\,d\tau \,\sin \omega (t - \tau )
\,q(\tau ) \,.
\end{eqnarray}
Here $Q_{i}(\omega ) \,, P_{i}(\omega )$ are initial environment
operators at $t = 0$, and $t_{f} (> t)$ is some arbitrary time.
The system evolution is then determined by the quantum Langevin
equation \cite{ford-lewis-oconnell},
\begin{eqnarray}
&&
\stackrel{..}{q} + V'(q) + 2\,\int_{0}^{t}\,d\tau \,\alpha _{I}(t - \tau )
\,q(\tau ) = F_{Q} \,, \label{quantum langevin eq} 
\\ &&
F_{Q}(t) = -\,\int_{\omega _{c}}^{\infty }\,d\omega \,\sqrt{r(\omega )}\,
\left( \,b_{i}(\omega )\,e^{-\,i\omega t} + 
b_{i}^{\dag }(\omega )\,e^{i\omega t}
\,\right) \,, 
\\ &&
\alpha _{I}(\tau ) = 
-\,\frac{i}{2}\,\int_{-\infty }^{\infty }\,
d\omega \,r(\omega )\,e^{-\,i\omega \tau } \,,
\label{dissipation kernel} 
\\ &&
r(\omega ) = \sum_{a}\,\frac{c_{a}^{2}}{2\omega _{a}}\,
\left( \, \delta (\omega - \omega _{a}) - \delta (\omega + \omega _{a})
\,\right) \nonumber 
\\ &&
\hspace*{1cm} 
= \int_{\omega _{c}}^{\infty }\,d\omega \,D(\omega )\,\frac{c^{2}(\omega )}
{2\omega } - \int_{-\infty }^{-\omega _{c}}\,
d\omega \,D(\omega )\,\frac{c^{2}(\omega )}{2\omega } \,.
\end{eqnarray}
$b_{i}(\omega )$ and $b_{i}^{\dag }(\omega )$ are the initial creation
and the annihilation operators of the environment part of variables.
$D(\omega )$ is the density of states per unit frequency.
The fundamental quantity $r(\omega )$
that characterizes both the environment spectrum and its coupling 
to the system is called the response weight.
The right hand side of eq.(\ref{quantum langevin eq}) describes the 
random force by the environment,
and it is characterized by the following correlation function, 
evaluated in the initial state of environment:
\begin{equation}
\langle \left( F_{Q}(\tau )F_{Q}(s)\right)_{{\rm sym}} \rangle_{{\rm env}}
= \int_{\omega _{c}}^{\infty }\,d\omega \,r(\omega )\,\cos \omega (\tau -s)
\,\langle 2n_{i}(\omega ) + 1 \rangle_{{\rm env}} \,.
\end{equation}
Only the symmetric part of the correlation is written here.
\( \:
n_{i}(\omega ) = b_{i}^{\dag }(\omega )b_{i}(\omega )
\: \)
is the number operator of the initial environmental variable.
For instance, in thermal state of temperature $T = 1/\beta $,
\begin{equation}
\langle\, 2n_{i}(\omega ) + 1\, \rangle_{{\rm env}} 
= \coth (\frac{\beta \omega }{2})
\,.
\end{equation}

To go further, we shall take dynamics of a single harmonic oscillator for
the small system,
\begin{eqnarray}
L_{q} = \frac{1}{2}\, (\frac{dq}{dt})^{2} - \frac{1}{2}\, \omega _{0}^{2}\,
q^{2} \,.
\end{eqnarray}
Even with this limitation there are an important
class of applications as we shall later discuss.
It turns out \cite{jmy-96-2} 
that the entire potential of the system plus the environment
can be diagonalized. Although it can be diagonalized via real orthogonal
matrix, it is more convenient to attach some physically unobservable
phase factors to this transformation.
The diagonal oscillator coordinates, $\tilde{Q}(\omega )$, along with
inverse relations, are then given by
\begin{eqnarray}
&& \hspace*{-1cm}
\tilde{Q}(\omega ) = Q(\omega ) - \sqrt{2\omega \,r(\omega )}
\,F(\omega - i0^{+})\,
\left( \,q - 
\int_{\omega _{c}}^{\infty }\,d\omega '\,\frac{\sqrt{2\omega '
\,r(\omega ')}}{\omega '\,^{2} - \omega ^{2} + i 0^{+}}\,Q(\omega ')
\,\right) \,, \label{diagonal oscillator} 
\\ && \hspace*{-1cm}
\tilde{P}(\omega ) = P(\omega ) - \sqrt{2\omega \,r(\omega )}\,
F(\omega - i0^{+} )\,\left( \,p - 
\int_{\omega _{c}}^{\infty }\,d\omega '\,\frac{\sqrt{2\omega '
\,r(\omega ')}}{\omega '\,^{2} - \omega ^{2} + i 0^{+}}\,P(\omega ')
\,\right) \,, 
\\ &&
q = -\,\int_{\omega _{c}}^{\infty }\,d\omega \,\sqrt{2\omega\, r(\omega )}\,
F^{*}(\omega - i0^{+} )\,\tilde{Q}(\omega ) \,, 
\\ &&
p = -\,\int_{\omega _{c}}^{\infty }\,d\omega \,\sqrt{2\omega\, r(\omega )}\,
F^{*}(\omega - i0^{+} )\,\tilde{P}(\omega ) \,, 
\\ &&
Q(\omega ) = \tilde{Q}(\omega ) + \sqrt{2\omega\, r(\omega )}\,
\int_{\omega _{c}}^{\infty }\,d\omega' \,
\frac{\sqrt{2\omega'\, r(\omega' )}\,F^{*}(\omega' - i0^{+} )}
{\omega ^{2} - \omega '\,^{2} - i0^{+}}\,\tilde{Q}(\omega' ) \,, 
\\ &&
P(\omega ) = \tilde{P}(\omega ) + \sqrt{2\omega\, r(\omega )}\,
\int_{\omega _{c}}^{\infty }\,d\omega' \,
\frac{\sqrt{2\omega'\, r(\omega' )}\,F^{*}(\omega' - i0^{+} )}
{\omega ^{2} - \omega '\,^{2} - i0^{+}}\,\tilde{P}(\omega' ) \,.
\end{eqnarray}
The analytic function that appears here is
\begin{eqnarray}
&&
F(z) = \frac{1}{-\,z^{2} + \omega _{0}^{2} + 2\pi 
\overline{G}(z)}
\,, \hspace{0.5cm} 
\overline{G}(z) = \int_{-\infty }^{\infty }\,\frac{d\omega }{2\pi }\,
\frac{r(\omega )}{z - \omega } \,.
\end{eqnarray}

This function has cuts along the real axis, 
\( \:
\omega > \omega _{c} 
\: \)
and 
\( \:
\omega < -\,\omega _{c} \,. 
\: \)
We assume the threshold $\omega _{c} > 0$ such that there is a gap between the
two cuts.
The following discontinuity relation holds:
\begin{eqnarray}
F(\omega + i 0^{+}) - F(\omega - i 0^{+}) = i
2\pi \,r(\omega )F(\omega + i 0^{+})F(\omega - i 0^{+})
\equiv i 2\pi H(\omega ) \,, 
\end{eqnarray}
along this cut.
An important property of $F(z)$ is that in the cut $z-$plane this function
is regular except on the real axis, and that it can be analytically
continued into the other Riemann sheets by the discontinuity formula.
In the second sheet $F(z)$ can be shown to have simple poles.
The location of these poles is given by
\begin{equation}
z ^{2} - \omega _{0}^{2} - 2\pi \,\overline{G}(z) 
+ 2\pi ir(z) = 0 \,. \label{spectrum zero} 
\end{equation}

The analytic property thus derived is based on the presence of the gap
at $-\,\omega _{c} < \omega < \omega _{c}$:
by analytic extention through this gap the boundary values at 
$\omega \pm i 0^{+}$ with 
\( \:
\, - \infty < \omega < - \omega _{c} \,
\: \)
and
\( \:
\omega _{c} < \omega < \infty \,
\: \)
become related.
Without the gap as in the case of $\omega _{c} = 0$, there is no unique
way to analytically extend $\overline{G}(z)$ (hence $F(z)$) with a given
$r(\omega )$ at $\omega \geq 0$.
Indeed, the gapless case is difficult to deal with in complete generality.
In the case of odd regular function for $r(z)$ 
the analytic extention is possible.
On the other hand, in cases with the spectrum weight behaving like
\( \:
r(\omega ) \propto \omega ^{\alpha } 
\: \)
(with $\alpha $ a positive non-integer)
near $\omega = 0$, the simple analytic extention in terms of the two
relevant Riemann sheets is impossible.

What happened to the subsystem spectrum?
The equation that governs the real part of its
energy eigenvalue squared $\lambda $ is given by
\begin{equation}
f(\lambda ) = \lambda \,, \hspace{0.5cm} 
f(\lambda ) \equiv \omega _{0}^{2} + {\cal P}\,\int_{\omega _{c}}^{\infty }\,
d\omega \,\frac{2\omega\, r(\omega )}{\lambda - \omega ^{2}} \,,
\label{spectum eq} 
\end{equation}
with ${\cal P}$ denoting the principal part of integration.
The function $f(\lambda )$ is related to the analytic function by
\begin{eqnarray}
f(\lambda ) - \lambda = \Re F^{-1}(\sqrt{\lambda } + i0^{+}) \,,
\end{eqnarray}
with $\Re $ denoting the real part.
In the $\lambda < \omega _{c}^{2}$ region one readily derives
from the monotonic behavior of $f(\lambda )$ 
the condition for absence of isolated spectrum at
\( \:
0 < \omega < \omega_{c}
\: \),
\begin{eqnarray}
f(\omega _{c}^{2 -}) > \omega _{c}^{2} \,, 
\label{condition of no real pole} 
\end{eqnarray}
with $^{-}$ indicating the limit from below.
Consistency further requires $\omega _{0} > \omega _{c}$.
When this condition (\ref{condition of no real pole}) is obeyed, 
the system frequency $\omega _{0}$ moves,
with increase of the coupling $c(\omega )$, into the second Riemann
sheet of the complex $\omega $ plane across the cut starting
at $\omega = \omega _{c}$.
It finally settles at the zero of eq.(\ref{spectrum zero}).
The imaginary part of this zero then gives the decay rate of 
any initial excitation of the subsystem, as will be made clear shortly.

When the response weight $r(\omega ) \rightarrow $ a constant 
as $\omega \rightarrow \infty $, as often happens and in field theory 
models later discussed in particular,
one has to subtract a term from integrals  containing the response weight.
This corresponds to renormalization of the bare frequency $\omega _{0}$ with
the frequency shift,
\begin{equation}
\delta \omega ^{2} = 
- \,2\,\int_{\omega _{c}}^{\infty }\,d\omega \,\frac{r(\omega )}{\omega} 
\,.
\end{equation}
The renormalized frequency is given by
\begin{equation}
\omega _{R}^{2} = \omega _{0}^{2} + \delta \omega ^{2} =
\omega _{0}^{2} - 2\,\int_{\omega _{c}}^{\infty }\,d\omega \frac{r(\omega )}
{\omega } \,. \label{renormalized frequency} 
\end{equation}
The eigenvalue equation written using $\omega _{R}^{2}$ is then modified to
\begin{eqnarray}
\lambda - \omega _{R}^{2} - \lambda \,{\cal P}\,
\int_{\omega _{c}}^{\infty }\,d\omega \,\frac{2r(\omega )}
{\omega\, (\,\lambda - \omega ^{2}\,)} = 0
\,. \label{spectrum eigenvalue} 
\end{eqnarray}
The condition of stability of the entire system requires that 
the smallest eigenvalue $\lambda > 0$, giving
\( \:
\omega _{R}^{2} > 0
\,,
\: \)
or equivalently
\begin{equation}
\omega _{0}^{2} > 2\,\int_{\omega _{c}}^{\infty }\,d\omega \,
\frac{r(\omega )}{\omega } \,, 
\end{equation}
if the integral on the right hand side is convergent.
In the discussion that follows the original subsystem frequency $\omega _{0}$
appears only through the renormalized $\omega _{R}$ or the real part of
the zero (\ref{spectrum zero}). In this sense $\omega _{0}$ is not
a fundamental parameter when physical consequences are discussed.

The important quantity for physical interpretation of the results
that follow is the overlap probability between the true eigen operator
$\tilde{Q}(\omega )$ with the original system variable $q$, 
which is calculable from eq.(\ref{diagonal oscillator}):
\begin{eqnarray}
|\,\langle \langle q\,|\,\tilde{Q}(\omega ) \rangle \rangle\,|^{2}
&=& 
2\omega\, r(\omega )\,|F(\omega - i0^{+})|^{2} = 2\omega \,H(\omega )
\nonumber \\
 &=&
\frac{2\omega\, r(\omega )}{(\,\omega ^{2} - \omega _{0}^{2}
- \Pi (\omega )\,)^{2} + (\pi r(\omega ))^{2}} \,.
\label{overlap probability} 
\end{eqnarray}
We decomposed the analytic $\overline{G}(\omega - i0^{+})$ into
the real and the imaginary parts,
\begin{eqnarray}
&&
2\pi \,\overline{G}(\omega - i0^{+}) = \Pi (\omega ) + i\pi \,
r(\omega ) \,,
\\ &&
\Pi (\omega ) = {\cal P}\,\int_{-\infty }^{\infty }\,d\omega '\,
\frac{r(\omega ')}{\omega - \omega '} \,.
\end{eqnarray}
In the weak coupling limit of $c(\omega ) \rightarrow 0$ the overlap
probability (\ref{overlap probability}) has the Breit-Wigner form
with a complex pole at the zero of eq.(\ref{spectrum zero}).
As will be more evident shortly, it physically means that
any system excitation is damped with a decay rate,
\begin{equation}
2\gamma  = \frac{\pi \,r(\bar{\omega })}{\bar{\omega }} \,, 
\end{equation}
with $\bar{\omega }$ the real part of the pole of $F(z)$.

With complete diagonalization it should be possible to explicitly write
the operator solution. The solution is
\begin{eqnarray}
&&
q(t) = -\,\int_{\omega _{c}}^{\infty }\,d\omega \,\sqrt{2\omega\,r(\omega )}\,
F^{*}(\omega - i 0^{+})\,\tilde{Q}(\omega \,, t) \,, 
\\ &&
\tilde{Q}(\omega \,, t) = \cos (\omega t)\,\tilde{Q}_{i}(\omega ) +
\frac{\sin (\omega t)}{\omega }\,\tilde{P}_{i}(\omega ) \,.
\end{eqnarray}
The initial values $\tilde{Q}_{i} \,, \tilde{P}_{i}$ are then rewritten
in terms of the original variables,
\( \:
q \,, p 
\: \)
and 
\begin{equation}
b_{i}(\omega ) = \sqrt{\frac{\omega }{2}}\,Q_{i}(\omega ) + 
\frac{i}{\sqrt{2\omega }}\,P_{i}(\omega ) 
\,, \hspace{0.5cm} b_{i}^{\dag } \,.
\end{equation}
After some straightforward calculation one finds \cite{jmy-96-2} that
\begin{eqnarray}
q(t) &=& p_{i}\,g(t) + q_{i}\,\dot{g}(t) \nonumber 
\\
&-& \int_{\omega _{c}}^{\infty }\,
d\omega \,\sqrt{r(\omega )}\,\left( 
\,h^{*}(\omega \,, t)\,e^{-\,i\omega t}\,b_{i}(\omega ) + 
h(\omega \,, t)\,e^{i\omega t}\,b_{i}^{\dag }(\omega ) \,\right) 
\,, \label{q-solution} 
\\
p(t) &=& 
p_{i}\,\dot{g}(t) + q_{i}\,\stackrel{..}{g}(t) 
\nonumber \\
&-& \int_{\omega _{c}}^{\infty }\,
d\omega \,\sqrt{r(\omega )}\,\left( 
\,k^{*}(\omega \,, t)\,e^{-\,i\omega t}\,b_{i}(\omega ) + 
k(\omega \,, t)\,e^{i\omega t}\,b_{i}^{\dag }(\omega )
\,\right) \,, \label{p-solution} 
\end{eqnarray}
where we introduced
\begin{eqnarray}
g(t) &=& 2\,\int_{\omega _{c}}^{\infty }\,d\omega 
\,H(\omega )\,\sin (\omega t)\,, \label{g-def} 
\\
h(\omega \,, t) &=& \int_{0}^{t}\,d\tau \,g(\tau )\,e^{-\,i\omega \tau } \,, 
\\
k(\omega \,, t) &=& \int_{0}^{t}\,d\omega \,\dot{g}(\tau )\,e^{-\,i
\omega \tau } = 
g(t)e^{- i\omega t} + i\omega h(\omega \,, t) \,.
\end{eqnarray}

These solutions, eqs.(\ref{q-solution}), (\ref{p-solution}), are
fundamental to any calculation of correlators such as
\( \:
\langle q(\tau )q(s) \rangle_{{\rm env}}
\: \).
Moreover, it completely solves the quantum Langevin equation,
 (\ref{quantum langevin eq}),
albeit for the limited case of system dynamics of the harmonic oscillator.
The operator solution consists of two parts, 
one that depends on the system initial
operator values, $q_{i} \,, p_{i}$, and the other that depends on
the initial environment values.
Thus the memory effect of initial system dependence is governed by
the function $g(t)$.

To see more on the memory effect, let us first ignore the random force
caused by environment, 
by taking the ensemble average over the initial environment variables
that are assumed to have no expectation value,
\( \:
\langle Q_{i}(\omega ) \rangle = \langle P_{i}(\omega ) \rangle = 0 \,.
\: \)
The expectation value of the system variable then follows \cite{jmy-96-2} 
\begin{eqnarray}
&&
\frac{d\,\langle p \rangle}{dt} = -\,\Omega^{2}(t)\,\langle q \rangle
- C(t)\,\langle p \rangle \,, \hspace{0.5cm} 
\frac{d\,\langle q \rangle}{dt} = \langle p \rangle \,,
\\ &&
\Omega ^{2}(t) = \frac{\dot{g}\stackrel{...}{g} - \stackrel{..}{g}^{2}}
{g\stackrel{..}{g} - \dot{g}^{2}} \,, \hspace{0.5cm} 
C(t) = \frac{\dot{g}\stackrel{..}{g} - g\stackrel{...}{g}}
{g\stackrel{..}{g} - \dot{g}^{2}} \,.
\label{time dependent friction} 
\end{eqnarray}
The quantities that appear in this equation are 
the time dependent friction ($C(t)$) and
the time dependent frequency squared ($\Omega^{2}(t)$), 
incorporating environmental effects. 

Actually the ensemble average is not necessary.
Using the explicit form of solution, one gets after eliminating
initial $q_{i} \,, p_{i}$ dependence,
\begin{eqnarray}
&&
\frac{d^{2}q}{dt^{2}} + \Omega ^{2}(t)\,q + C(t)\,\frac{dq}{dt} =
-\,\Omega ^{2}(t)\,f_{q} - C(t)\,f_{p} - \dot{f}_{p} \,, 
\\ &&
f_{q} =
\int_{\omega _{c}}^{\infty }\,
d\omega \,\sqrt{r(\omega )}\,\left( 
\,h^{*}(\omega \,, t)\,e^{-\,i\omega t}\,b_{i}(\omega ) + ({\rm h.c.})
\,\right) \,, 
\\ &&
f_{p} =
\int_{\omega _{c}}^{\infty }\,
d\omega \,\sqrt{r(\omega )}\,\left( 
\,k^{*}(\omega \,, t)\,e^{-\,i\omega t}\,b_{i}(\omega ) + ({\rm h.c.})
\,\right) \,.
\end{eqnarray}
This may be viewed as a local form of quantum Langevin equation:
the operator acting on the system variable $q$ is the local
differential operator with local coefficient functions of time.
Note that this form is not available in the general case of
arbitrary potential.
In the general case the integro-differential equation is available.

Let us note that the function $g(t)$ has the behavior of damped
oscillation,
\begin{equation}
g(t) \sim \frac{1}{\bar{\omega }}\,\sin (\bar{\omega }t)\,e^{-\,\gamma t}
\,, 
\end{equation}
if one takes the Breit-Wigner form for $H(\omega )$, or the pole
approximation for $F(z)$.
When the coupling is weak, this pole approximation is excellent
in many practical applications.
In this case
\begin{equation}
\Omega ^{2}(t) = \bar{\omega }^{2} + \gamma ^{2} 
\equiv \Omega _{0}^{2} \,, \hspace{0.5cm} 
C(t) = 2\gamma \,.
\end{equation}
Thus the effect of the environment is very simple in the pole
approximation:
aside from the random fluctuating environmental force, the system experiences
the frequency renormalization and a constant friction:
\begin{equation}
\frac{d^{2}\langle q \rangle}{dt^{2}} + 2\gamma \frac{d\langle q \rangle}{dt}
+ \Omega _{0}^{2}\,\langle q \rangle = 0 \,.
\end{equation}

There are however physical phenomena that the pole approximation fails
to describe, as will be shown later.
Let us simply state here that both early time and late time behavior
of the system variables are not correctly described by the pole 
appproximation.
For instance, the correct $C(t) = O[t^{3}]$ as $t \rightarrow 0^{+}$ and
at late times both of $\Omega ^{2}(t)$ and $C(t)$ decrease with
powers of time.
These are not obeyed in the pole approximation.
Indeed, the pole approximation violates the positivity of the density matrix
at early times, as will be shown later.

The correct final behavior at $t \rightarrow \infty $ is the power law
of time, as seen in the following way.
Using the discontinuity formula, one may rewrite the $\omega $ integration
for $g(t)$, eq.(\ref{g-def}), containing 
\( \:
H(\omega ) = \left( \,F(\omega + i 0^{+}) - F(\omega - i 0^{+})\,\right)
/ (2\pi i)
\: \)
along the real axis into the $F(z)$ integration, the
complex $z$ running both slightly above and below the cuts.
The factor $\sin (\omega t)$ is replaced by $\Im e^{-\,i\omega t}$
in this procedure.
A half of this complex contour can be deformed into the second sheet,
and one thereby encounters simples poles in the second sheet.
The intregral for $g(t)$ may then be expressed as 
the sum of the pole contribution (at $z = z_{0} $ with 
$ \Im z_{0} < 0$ for a single pole) in the second sheet and the contribution
parallel to the imaginary axis passing through $z = \omega _{c} $, both
in the first (\lromn 1) and in the second (\lromn 2)
sheet \cite{goldberger-watson}:
\begin{eqnarray}
&& \hspace*{-2cm}
g(\tau ) =
\Im \left( Ke^{-i\Re z_{0}\tau } \right)\,
e^{\Im z_{0}\tau } 
+
\Im \left[ \frac{e^{i\omega _{c}\tau }}{\pi }\int_{0}^{\infty }dy\,
e^{- y\tau }\,\left( \,F_{{\rm \lromn 1}}(\omega _{c} + iy) - 
F_{{\rm \lromn 2}}(\omega _{c} + iy)\,\right)\, \right] \,, \label{g-integral} 
\end{eqnarray}
with
\( \:
K^{-1} = z_{0} - \pi \overline{G}'(z_{0} ) + i\pi r'(z_{0})\,.
\: \)

As seen from this formula, 
the pole contribution given by the first term describes
the exponential decay as mentioned already.
The rest of contribution gives the power law decay at very late times;
$\propto t^{-\alpha -1}$ \cite{jmy-96-2}.
The power $-\,\alpha -1$ is related to 
the threshold behavior of the response weight,
\( \:
r(\omega ) \:\propto  \: (\omega - \omega _{c})^{\alpha } \,.
\: \)
This continuous integral also describes the correct early time
behavior.

Emergence of the power law term in the quantum Brownian motion
has been noted in some specific
models,
but we find this behavior as a general property 
in the presence of the non-local dissipation.
We would like to stress that the analytic structure of the cut
$\omega $ plane is very important to derive the power law decay.
For instance, if one takes the Ohmic form of the response weight given
by \cite{hjmy 96} 
\begin{eqnarray}
r^{(4)}(\omega ) = \frac{4c\Omega \gamma\, \omega }
{(\omega^{2} - \Omega ^{2}
+ \frac{\gamma ^{2}}{4})^{2} + \Omega ^{2}\gamma ^{2}} \,, 
\end{eqnarray}
then one does not obtain the power law decay.
The point is that this function,  consisting of pole terms alone, 
does not have the branch point singularity needed for the power law
behavior.

Quadratic quantities averaged over the initial environment ensemble
are computed as \cite{jmy-96-2} 
\begin{eqnarray}
&& 
\langle\, q^{2}(t)\, \rangle = 
\int_{\omega _{c}}^{\infty }\,d\omega \,
\langle\, 2n_{i}(\omega ) + 1\, \rangle_{{\rm env}}
\,r(\omega )\,|h(\omega \,, t)|^{2} 
\nonumber \\ &&
\hspace*{1cm} 
+ \,
g^{2}(t)\,\langle\, p_{i}^{2}\, \rangle + 
\dot{g}^{2}(t)\,\langle\, q_{i}^{2}\, \rangle + g(t)\dot{g}(t)\,
\langle\, p_{i}q_{i} + q_{i}p_{i}\, \rangle \,, 
\\ && 
\langle\, p^{2}(t)\, \rangle =
\int_{\omega _{c}}^{\infty }\,d\omega \,
\langle\, 2n_{i}(\omega ) + 1\, \rangle_{{\rm env}}
\,r(\omega )\,|k(\omega \,, t)|^{2} 
\nonumber 
\\ &&
\hspace*{1cm} 
+\,
\dot{g}^{2}(t)\,\langle\, p_{i}^{2}\, \rangle 
+ \stackrel{..}{g}^{2}(t) \langle\, q_{i}^{2}\, \rangle + 
\dot{g}(t)\stackrel{..}{g}(t)
\langle\, p_{i}q_{i} + q_{i}p_{i}\, \rangle \,, 
\\ &&
\frac{1}{2}\, \langle\, q(t)p(t) + p(t)q(t)\, \rangle
= 
\int_{\omega _{c}}^{\infty }\,d\omega \,
\langle\, 2n_{i}(\omega ) + 1\, \rangle_{{\rm env}}
\,r(\omega )\,h(\omega \,, t)
k^{*}(\omega \,, t) 
\nonumber \\ &&
\hspace*{0.5cm} +\,
\dot{g}(t)g(t)\,\langle\, p_{i}^{2}\, \rangle + \dot{g}(t)\stackrel{..}{g}(t)
\langle\, q_{i}^{2}\, \rangle + (\,\dot{g}^{2}(t) + g(t)\stackrel{..}{g}(t)\,)
\,\frac{1}{2}\, \langle\, p_{i}q_{i} + q_{i}p_{i}\, \rangle \,.
\end{eqnarray}
Here we assumed both
\( \:
\langle b_{i}(\omega ) \rangle = \langle b_{i}^{\dag }(\omega ) \rangle
= 0
\: \)
and
\( \:
\langle b_{i}(\omega )b_{j}(\omega ') \rangle =
\langle b_{i}^{\dag }(\omega )b_{j}^{\dag }(\omega ') \rangle = 0 \,.
\: \)
Equivalent expressions are obtained using identities:
\begin{eqnarray}
&& \hspace*{-1cm}
\int_{\omega _{c}}^{\infty }\,d\omega \,\coth (\frac{\beta \omega }{2})\,
r(\omega )\,|h(\omega \,, t)|^{2}
=
2\,\int_{0}^{t}\,d\tau \,\int_{0}^{\tau }\,ds\,
g(t - \tau )\alpha _{R}(\tau - s)g(t - s)
\,, \\ && \hspace*{-1cm}
\int_{\omega _{c}}^{\infty }\,d\omega \,\coth (\frac{\beta \omega }{2})\,
r(\omega )\,|k(\omega \,, t)|^{2}
=
2\,\int_{0}^{t}\,d\tau \,\int_{0}^{\tau }\,ds\,
\dot{g}(t - \tau )\alpha _{R}(\tau - s)\dot{g}(t - s)
\,, \\ && \hspace*{-1cm}
\alpha _{R}(\tau ) = \int_{\omega _{c}}^{\infty }\,d\omega \,
\langle\, 2n_{i}(\omega ) + 1\, \rangle\,\frac{\cos (\omega \tau )}{2\omega }\,.\end{eqnarray}

Let us turn to the late time behavior.
At asymptotically late times the initial value dependence disappears,
as mentioned above, leading to
\begin{eqnarray}
&&
q(t) \:\rightarrow  \: -\,\int_{\omega _{c}}^{\infty }\,d\omega \,
\sqrt{r(\omega )}\,\left( \,F^{*}(\omega - i 0^{+})\,e^{-\,i\omega t}
\,b_{i}(\omega ) + ({\rm h.c.})\,\right) \,, 
\\ &&
p(t) \:\rightarrow  \: i\,\int_{\omega _{c}}^{\infty }\,d\omega \,\omega \,
\sqrt{r(\omega )}\,\left( \,F^{*}(\omega - i 0^{+})\,e^{-\,i\omega t}
\,b_{i}(\omega ) - ({\rm h.c.})\,\right) \,,
\end{eqnarray}
since $h(\omega \,, \infty ) = F(\omega - i0^{+})$.
The system evolution is thus governed by the initial distribution of
the environment variables, $b_{i}(\omega ) \,, b_{i}^{\dag }(\omega )$,
with the following probability functions;
\begin{eqnarray}
&&
r(\omega )|h(\omega \,, \infty )|^{2} = H(\omega ) = 
\frac{|\,\langle\langle \, q|\,\tilde{Q}(\omega )\rangle  \rangle\,|^{2}}
{2\omega } \,, 
\hspace{0.5cm}  {\rm for} \; q(t) \,,
\\ &&
r(\omega )|k(\omega \,, \infty )|^{2} = \omega^{2}H(\omega ) = 
\frac{\omega }{2}\, 
|\,\langle\langle \, q|\,\tilde{Q}(\omega )\rangle  \rangle\,|^{2} 
\,, \hspace{0.5cm}  {\rm for} \; p(t) \,.
\end{eqnarray}
Note that the overlap probability
\( \:
|\,\langle\langle \, q|\,\tilde{Q}(\omega )\rangle  \rangle\,|^{2} 
\: \)
has a universal character,
being determined by general properties of the system and the environment alone,
irrespective of their particular initial states.
The overlap probability between the system $q$ variable and the diagonal
$\tilde{Q}(\omega )$ respects the unitarity relation,
\begin{equation}
\int_{\omega _{c}}^{\infty }\,d\omega \,
|\,\langle\langle \, q|\,\tilde{Q}(\omega )\rangle  \rangle\,|^{2}
= 
\int_{\omega _{c}}^{\infty }\,d\omega \,2\omega \,H(\omega ) = 1\,.
\end{equation}

Asymptotic values of averaged quadratic quantities are
\begin{eqnarray}
\langle \,q^{2}(\infty )\, \rangle &=&
\int_{\omega _{c}}^{\infty }\,d\omega \,
\langle \,2n_{i}(\omega ) + 1 \, \rangle_{{\rm env}}\,H(\omega )
\,, \\
\langle \,p^{2}(\infty )\, \rangle &=&
\int_{\omega _{c}}^{\infty }\,d\omega \,
\langle \,2n_{i}(\omega ) + 1 \, \rangle_{{\rm env}}\,\omega ^{2}\,H(\omega )
\,.
\end{eqnarray}
Taking the thermal bath with
\( \:
\langle \,2n_{i}(\omega ) + 1 \, \rangle_{{\rm env}} 
= \coth (\frac{\beta \omega }{2})
\,, 
\: \)
along with the pole approximation, this gives the well known results:
\begin{equation}
\langle \,q^{2}(\infty )\, \rangle \approx \frac{T}{\bar{\omega }^{2}}
\,, \hspace{0.5cm} 
\langle \,p^{2}(\infty )\, \rangle \approx T \,.
\end{equation}

Let us discuss more closely the asymptotic form of the occupation number
in thermal bath,
\begin{eqnarray}
n(\infty ) &\equiv& \frac{1}{2}\, \langle\, \frac{p^{2}(\infty )}
{\bar{\omega}} 
+ \bar{\omega}\,q^{2}(\infty )\, \rangle - \frac{1}{2} 
\nonumber \\ 
&=& 
\frac{1}{2}\, \int_{\omega _{c}}^{\infty }\,d\omega \,\coth (\frac{\beta 
\omega }{2})\,(\bar{\omega} + \frac{\omega ^{2}}{\bar{\omega}})\,H(\omega )
- \frac{1}{2} \,. 
\end{eqnarray}
The temperature dependent part of the occupation number is then
\begin{equation}
n^{\beta  } = 
\int_{\omega _{c}}^{\infty }\,d\omega \,\frac{1}{e^{\beta \omega }
- 1}\,(\bar{\omega}  + \frac{\omega ^{2}}{\bar{\omega} })\,H(\omega ) \,.
\end{equation}
When the pole term dominates, or equivalently one approximates $H(\omega )$
by the Breit-Wigner function,
$n^{\beta }$ has the factor of Boltzmann suppression,
\( \:
e^{-\,\bar{\omega }/T}
\: \)
at low temperatures.
But the pole approximation is not good at low temperatures.
Indeed, let us examine a typical example by taking the form of
\( \:
r(\omega ) = c\,(\omega - \omega _{c})^{\alpha } \,, 
\: \)
in the range of $\omega _{c} < \omega < \Omega $
($\Omega \gg \omega _{c}$). We assume $0 < \alpha < 1$  and 
\( \:
\bar{\omega } \gg  {\rm Max}\;(\,\omega_{c} \,, T\,)
\: \).
The result is 
\begin{equation}
n^{\beta  } \approx \frac{c}{\bar{\omega }^{3}}\,\Gamma (\alpha + 1)\,
e^{-\,\beta \omega _{c}}\,T^{\alpha + 1} \,.
\end{equation}
$\Gamma $ is the Euler's gamma function.

This shows that instead of the exponential suppression at low temperatures,
what is left in medium after the decay has a power-law 
behavior of temperature dependence ($\propto T^{\alpha + 1}$).
An implication of this behavior to the unstable particle decay
will be discussed in the next section.
It means that the remnant fraction in thermal medium does not suffer from
the Boltzmann suppression factor at temperatures even much lower than
the mass of unstable particle.

Related time evolution is as follows.
The corresponding behavior of $g(t)$ computed from the continuous
part of $H(\omega )$ integration gives
\begin{eqnarray}
g(t) \approx -\,\frac{2c}{\bar{\omega }^{4}}\,\Gamma (\alpha + 1)\,
\frac{\cos (\,\omega _{c}\,t + \frac{\pi }{2}\,\alpha \,)}
{t^{\alpha + 1}} \,.
\end{eqnarray}
One can estimate the transient time $t_{*}$ from the exponential period to
the power period by equating the two formulas of $g(t)$ in their
respective ranges, to obtain
\begin{equation}
t_{*} \approx \frac{1}{\gamma }\,\ln \left( \frac{\bar{\omega }^{3}}
{2c\,\Gamma(\alpha + 1)\,\gamma ^{\alpha + 1}}\right)
 \,,
\end{equation}
with 
\( \:
\gamma = -\,\Im z_{0} \,
\: \)
the decay rate.
For a very small $c$ the factor inside the logarithm becomes large
($\propto c^{-\,2\alpha - 3}$), and by the time $t_{*}$ the initial
population has decreased like
\begin{equation}
e^{-\,2\gamma t_{*}} \:\propto  \: c^{4\alpha + 6} \,.
\end{equation}
It may thus be claimed that the power law behavior is difficult
to observe.
But we shall later discuss that this may not be so in cosmology.

We shall refer to our original paper \cite{jmy-96-2} 
on detailed discussion of correlation functions.
But let me mention the asymptotic form.
For both $t_{1}$ and $t_{2}$ in the asymptotic late
time region,
\begin{eqnarray}
&&
\langle \,q(t_{1})q(t_{2})\, \rangle \:\rightarrow  \:
\int_{\omega _{c}}^{\infty }\,d\omega 
H(\omega )\,
\cos \omega (t_{1} - t_{2})\,\coth (\frac{\beta \omega }{2})
\,, 
\\ &&
\langle\, p(t_{1})p(t_{2})\, \rangle \:\rightarrow  \:
\int_{\omega _{c}}^{\infty }\,d\omega\,
\omega ^{2}\,H(\omega )\,
\cos \omega (t_{1} - t_{2})\,\coth (\frac{\beta \omega }{2}) \,.
\end{eqnarray}
The other correlator vanishes:
\( \:
\langle \,q(t_{1})p(t_{2}) + p(t_{2})q(t_{1})\, \rangle \rightarrow 0 \,.
\: \)

Finally let us mention how some of our fundamental formulas
are related to quantities in complete thermal equilibrium.
The limiting values of the analytic function $\overline{G}(z)$ as
$z \:\rightarrow  \: $ real $\omega $ are related to
the well-known real-time thermal Green's function \cite{fetter-walecka}, 
when one takes the initial environment in thermal state. For a single
oscillator,
\begin{eqnarray}
&&
G^{R}(\omega ) \equiv  \overline{G}(\omega + i\epsilon ) \,, \hspace{0.5cm} 
G^{A}(\omega ) \equiv  \overline{G}(\omega - i\epsilon ) \,,
\\ &&
\hspace*{-2cm}
\frac{1}{1 - e^{-\beta \omega }}\,G^{R}(\omega ) +
\frac{1}{1 - e^{\beta \omega }}\,G^{A}(\omega ) 
=
i\,\int_{-\infty }^{\infty }\,d\tau \,e^{i\omega \tau }\,
{\rm tr}\; \left( \,\rho _{\beta }\,T\left[ \,
Q(\tau )\,Q(0)\,\right]\,\right) \,.
\end{eqnarray}

\vspace{0.5cm} 
The operator method is thus very powerful.
On the other hand, it is difficult to extract quantum statistical nature of
the subsystem state in this approach.
For this purpose, the path integral approach is useful, to which
we shall now turn.

\vspace{1cm} 
{\bf \lromn 2B Path integral method}

\vspace{0.5cm} 
The basic idea \cite{feynman-vernon} is that one is interested in the
behavior of the $q-$system alone
and traces out the environment variable altogether in the path
integral formula.
In the influence functional approach by Feynman and Vernon 
one directly deals with the probability instead of the amplitude.
This way one can compute the reduced density matrix 
that describes the state of the small system incorporating effects
of interacting environment.
We define the influence functional by convoluting with the initial
state of the environment. To do so we assume for technical reasons
that initially we may take
an environment state uncorrelated with the system.
The influence functional is thus obtained after integrating 
out the environment variables:
\begin{eqnarray}
\hspace*{-0.5cm}
&&
{\cal F}[\,q(\tau )\,, q'(\tau )\,] \equiv 
\int\,{\cal D}Q(\tau )\,\int\,{\cal D}Q'(\tau )\,\int\,dQ_{i}\,\int\,dQ'_{i}
\int\,dQ_{f}\,\int\,dQ'_{f}\,\delta (Q_{f} - Q'_{f})\,
\nonumber \\
&& \hspace*{1cm} 
\cdot K \left(\,q(\tau )\,,Q(\tau ) \,\right)\,
K^{*} \left( \,q'(\tau )\,,Q'(\tau ) \,\right)\,\rho_{i} (Q_{i}\,, Q'_{i}) \,,
\\
&&
K \left( \,q(\tau )\,,Q(\tau ) \,\right) =
\exp \left( \,iS_{0}[Q] + iS_{{\rm int}}[q \,, Q]\,\right) \,, \\
&&
S_{0}[Q] + S_{{\rm int}}[q \,, Q] = \int_{0}^{t}\,d\tau \,
\left( \,L_{Q}[Q] + L _{{\rm int}}[q\,, Q]\,\right) \,.
\end{eqnarray}
The influence functional is a functional of the entire path 
of the system $q(\tau )$ and its conjugate path $q'(\tau )$.
\begin{equation}
\rho_{i} (Q_{i}\,, Q'_{i}) = \sum_{n}\,w_{n}\,\psi _{n}^{*}(Q_{i}\,')
\psi _{n}(Q_{i}) \,,  \hspace{0.5cm} (0 \leq w_{n} \leq 1)
\end{equation}
is the initial density matrix of the environment, which can be any
mixture of pure quantum states with probability $w_{n}$.
What deserves to be stressed is that one does not observe the final
state of the environment, hence integration with respect to the final
values of $Q_{f} = Q'_{f}$ is performed here.

Once the influence functional is known, one may compute the transition
probability and any physical quantities of the $q-$system
by convoluting dynamics of the system under study. 
For instance, the transition probability is given, with introduction
of the density matrix $\rho ^{(R)}$, by
\begin{eqnarray}
&& 
\int\,dq_{f}\,\int\,dq'_{f}\,\psi_{f} ^{*}(q_{f})
\rho ^{(R)}(q_{f} \,, q'_{f})\,\psi_{f} (q'_{f}) \,, 
\\ && \hspace*{-1.5cm}
\rho ^{(R)} = 
\int\,{\cal D}q(\tau )\,\int\,{\cal D}q'(\tau )\,
\int\,dq_{i}\,\int\,dq'_{i}\,
\psi_{i} ^{*}(q'_{i})\,\psi_{i} (q_{i})
\,{\cal F}[\,q(\tau )\,, q'(\tau )\,]\,
e^{iS_{q}[q] - iS_{q}[q']} \,, 
\end{eqnarray}
where $\psi_{i\,, f} $'s are wave functions of the initial and
the final $q-$states, and $S_{q}[q]$ is the action of the $q-$system.

The form of the influence functional is dictated by general principles
such as probability conservation and causality. Feynman and Vernon 
found a closed quadratic form consistent with these,
\begin{eqnarray}
{\cal F}[\,q(\tau )\,, q'(\tau )\,] &=& \nonumber \\
&& \hspace*{-3cm}
\exp \left[\,-\,\int_{0}^{t }\,d\tau \,\int_{0}^{\tau }\,ds\,
\left( \,\xi (\tau )\alpha_{R}(\tau - s)\xi (s) + i\,\xi (\tau )
\alpha _{I}(\tau - s)X(s)\,\right)\,\right] \,, 
\label{influence-f def} \\
&& \hspace*{-2cm}
{\rm with} \;
\xi (\tau ) = q(\tau ) - q'(\tau ) \,, \hspace{0.5cm} 
X(\tau ) = q(\tau ) + q'(\tau ) \,.
\end{eqnarray}
Thus two real functions $\alpha _{i}(\tau )$ are all we need to
characterize the system-environment interaction.
These are defined here in the range of 
\( \:
\tau \geq 0 \,.
\: \)
The fact that $\alpha_{i} $ depends on the difference of time
variables, $\tau - s$, is due to the assumed stationarity of the environment.
The Feyman-Vernon formula is valid for general $L_{Q}[Q]$ and
$L_{q}[q]$, not limited to the harmonic oscillator model if the interaction
$L_{{\rm int}}[q \,, Q]$ is bilinear.

The correlation kernels appear in the influence functional
as a form of the nonlocal interaction and they
are the dissipation $\alpha _{I}$ and the noise $\alpha _{R}$. 
The dissipation kernel $\alpha _{I}$ thus computed
agrees with the one defined in eq.(\ref{dissipation kernel}) \cite{hjmy 96}.
Let us now specialize to the case of the oscillator bath of temperature
$T = 1/\beta $,
which is described for a single oscillator of frequency $\omega $ by
\begin{eqnarray}
\rho _{\beta }(Q \,, Q') &=& \left( \frac{\omega }{\pi \,\coth (\beta \omega 
/2)}\right)^{1/2}\,
\nonumber \\ &&
\cdot \exp \left[ \,-\,\frac{\omega }{2 \sinh (\beta \omega )}
\,\left( \,(Q^{2} + Q'\,^{2})\,\cosh (\beta \omega ) - 2 Q Q'\,\right)\,\right]
\,. \label{thermal density matrix} 
\end{eqnarray}
The noise kernel, along with the dissipation kernel, are then given by
\begin{eqnarray}
\alpha _{R}(\tau ) &=& \frac{1}{2}\, \int_{-\infty }^{\infty }\,d\omega \,
\coth (\frac{\beta \omega }{2})\,r(\omega )\,e^{-\,i\omega \tau } \,,
\\
\alpha _{I}(\tau ) &=& -\,\frac{i}{2}\,\int_{-\infty }^{\infty }\,
d\omega \,r(\omega )\,e^{-\,i\omega \tau } \,.
\end{eqnarray}
Combined together, it gives the real-time thermal Green's function:
\begin{eqnarray}
&& \hspace*{-0.5cm}
\alpha (\tau ) \equiv \alpha _{R}(\tau ) + i\alpha _{I}(\tau )
= 
\sum_{k}\,c_{k}^{2}\,{\rm tr}\; \left( \,\rho _{\beta }\,T\left[ \,
Q(\omega _{k} \,, \tau )\,Q(\omega _{k} \,, 0)\,\right]\,\right) \,, 
\\ && \hspace*{-0.5cm}
\alpha (\omega ) \equiv \int_{-\infty }^{\infty }\,d\tau \,\alpha (\tau )
\,e^{i\omega \tau } =
i\,\sum_{k}\,c_{k}^{2}\,\left( \frac{1}{\omega^{2} -
\omega_{k}^{2} + i\epsilon } - \frac{2\pi i}{e^{\beta \omega _{k}} - 1}
\,\delta (\omega^{2} - \omega _{k}^{2}) \right) \,.
\end{eqnarray}
As noted already, these are given in terms of the response weight
$r(\omega )$, and are governed by the analytic function $\overline{G}(z)$.

For the system dynamics we further assume a single harmonic oscillator
of frequency $\omega _{0}$.
In the path integral approach integration over the sum variable
$X(\tau )$ is trivial in this case,
since both the local part and the nonlocal action above
are linear in this variable:
\begin{eqnarray}
&&
\frac{1}{2}\, \int_{0}^{t}\,\left( \,\dot{\xi }(\tau )\dot{X}(\tau )
- \omega _{0}^{2}\,\xi (\tau )X(\tau )\,\right)  \nonumber 
\\ && \hspace*{-1cm}
- \,\int_{0}^{t }\,d\tau \,\int_{0}^{\tau }\,ds\,
\left( \,\xi (\tau )\alpha_{R}(\tau - s)\xi (s) + i\,\xi (\tau )
\alpha _{I}(\tau - s)X(s) \,\right) \,. 
\end{eqnarray}
Thus result of the path integration
of the system variable $X(\tau )$ gives the classical 
integro-differential equation for $\xi (\tau )$:
\begin{equation}
\frac{d^{2}\xi }{d\tau ^{2}} + \omega^{2} _{0}\,\xi (\tau ) +
2\,\int_{\tau }^{t}\,ds \,\xi (s)\,\alpha _{I}(s - \tau) = 0 \,.
\label{xi integro-eq} 
\end{equation}
The end result of the $\xi $ path integral then 
contains an integral of the form,
\begin{eqnarray}
-\,
\int_{0}^{t}\,d\tau \,\int_{0}^{\tau }\,ds\,\xi (\tau )\alpha _{R}(\tau - s)
\xi (s) \,,
\end{eqnarray}
using the classical solution $\xi (\tau )$ with specified boundary
conditions,
\( \:
\xi (0) = \xi _{i} \,, \hspace{0.3cm} \xi (t) = \xi _{f}
\: \).

In the local approximation often used the dissipation kernel 
is taken to have the form of
\begin{eqnarray}
\alpha _{I}(\tau ) = \delta \omega ^{2}\,\delta (\tau ) 
+ \eta \,\delta '(\tau ) \,, 
\end{eqnarray}
with $\delta \omega ^{2}$ representing 
the frequency shift and the $\eta $ term the local friction.
This choice enables one to solve the $\xi $ equation (\ref{xi integro-eq}) 
by elementary means.
On the other hand, the noise kernel is usually given by the response weight of
the form,
\begin{equation}
r(\omega ) = \frac{\eta }{\pi }\,\omega f(\frac{\omega }{\Omega }) \,, 
\end{equation}
with $f(x)$ some cutoff function and $\Omega $ a high frequency cutoff.
The cutoff is needed to tame the high frequency integral of 
$\alpha _{R}(t)$.
The simplest cutoff function 
\( \:
f(x) = \theta (1 - x)
\: \)
gives an approximate form of $\alpha _{I}(t)$ with the friction $\eta $
and
\begin{equation}
\delta \omega ^{2} \approx   -\,\frac{2}{\pi }\,\eta\,\Omega \,.
\end{equation}

At high temperatures this approximation 
reduces to the well known classical form of the fluctuation,
\begin{equation}
\alpha _{R}(\tau ) = \frac{\eta T}{\pi }\,\delta (\tau ) \,.
\end{equation}
At low temperatures, howerver,
the use of the cutoff function only in the noise
kernel while retaining the local form of the dissipation
makes validity of this approximation dubious.
Our path integral approach \cite{jmy-96-1} does not make this 
local approximation, and
instead uses exact solutions of the classical equation (\ref{xi integro-eq}).
This equation has been used in other approaches \cite{qbm path review}, too.

The rest of deduction uses the Laplace transform, and we shall be brief,
leaving technical details to our original paper \cite{jmy-96-1}.
Solution of the integro-differential equation (\ref{xi integro-eq}) 
is, using $g(\tau )$ defined by eq.(\ref{g-def}), given as
\begin{eqnarray}
\xi (\tau ) =
\xi _{i}\,\frac{g(t - \tau )}{g(t)} + \xi _{f}\,
\left( \,\dot{g}(t - \tau ) - \frac{g(t - \tau )\dot{g}(t)}{g(t)} \,\right)
\,, 
\end{eqnarray}
with the dot denoting derivative.
Both $g(t)$ and $\dot{g}(t)$ can be shown to satisfy the integro-differential
equation of the form ($x = g \;$ or $\dot{g}$),
\begin{eqnarray}
\frac{d^{2}x}{dt^{2}} + \omega _{0}^{2}\,x + 2\,\int_{0}^{t}\,d\tau \,
\alpha _{I}(t - \tau  )\,x(\tau ) = 0 \,.
\end{eqnarray}

The reduced density matrix of the quantum system at any time is obtained from
the action written in terms of the boundary values,
\( \:
S_{cl}(\xi _{f} \,, X_{f} \, ; \, \xi _{i} \,, X_{i}) \,, 
\: \)
by convoluting with the initial density matrix of the thermal environment.
This action is computed as
\begin{eqnarray}
i\,S_{{\rm cl}} &=& -\,\frac{U}{2}\,\xi _{f}^{2} - \frac{V}{2}\,\xi _{i}^{2}
- W\,\xi _{i}\,\xi _{f}  + 
\frac{i}{2}\,X_{f}\,\dot{\xi }_{f} - 
\frac{i}{2}\,X_{i}\,\dot{\xi }_{i} \,, 
\\
U &=& 2\,\int_{0}^{t }\,d\tau \,\int_{0}^{\tau }\,ds\,
z(\tau )\,\alpha _{R}(\tau - s)\,z(s) \,, 
\\
V &=& 2\,\int_{0}^{t }\,d\tau \,\int_{0}^{\tau }\,ds\,
y(\tau )\,\alpha _{R}(\tau - s)\,y(s) \,, 
\\
W &=& \int_{0}^{t }\,d\tau \,\int_{0}^{\tau }\,ds\,
\left( \,y(\tau )z(s) + y(s)z(\tau )\,\right)\,\alpha _{R}(\tau - s) \,, 
\\
y(\tau ) &=&
\frac{g(t - \tau )}{g(t)} \,, 
\\
z(\tau ) &=& \dot{g}(t - \tau ) - g(t - \tau )\frac{\dot{g}(t)}{g(t)} \,, 
\\
\dot{\xi }(\tau ) &=&
-\,\xi _{i}\frac{\dot{g}(t - \tau )}{g(t)} - \xi _{f}\,
\left( \,\stackrel{..}{g}(t - \tau ) - \frac{\dot{g}(t - \tau )\dot{g}(t)}
{g(t)}\,\right) \,.
\end{eqnarray}

The master equation for the reduced density matrix, which holds for
any initial state of the system, 
may be derived  \cite{lindblad} from this effective action:
\begin{eqnarray}
i\frac{\partial \rho }{\partial t} &=&
-\,2\frac{\partial ^{2}\rho }{\partial \xi \partial X} 
+ \frac{\Omega ^{2}(t)}{2}\,\xi X\rho - 
iC(t)\,\xi \frac{\partial \rho }{\partial \xi }
+ 2i\xi ^{2}\,D_{pp}(t)\rho 
+ 4D_{xp}(t)\,\xi \frac{\partial \rho }{\partial X} \,, \nonumber \\
\end{eqnarray}
where some new quantites are defined as
\begin{eqnarray}
D_{pp}(t) &=&
\frac{1}{2}\, \left( \,\frac{g\stackrel{...}{g} - \dot{g}\stackrel{..}{g}}
{g\stackrel{..}{g} - \dot{g}^{2}}\,U
+ \frac{\dot{g}}{2g}\,\frac{g^{2}\stackrel{...}{g} - 
2g\dot{g}\stackrel{..}{g} + \dot{g}^{3}}{g\stackrel{..}{g} - \dot{g}^{2}}
\,W - \frac{\dot{U}}{2} - \dot{g}\dot{W}\,\right) \,,
\\
D_{xp}(t) &=&
U - g\dot{W} + \frac{g^{2}\stackrel{...}{g} - 
2g\dot{g}\stackrel{..}{g} + \dot{g}^{3}}{g\stackrel{..}{g} - \dot{g}^{2}}\,W
\,.
\end{eqnarray}
The same master equation was derived by Hu, Paz, and Zhang
\cite{q-dissipation path}, and their derivation was later simplified by
ref \cite{halliwell-yu}.
These works however do not give exact solutions like ours.
Moreover, our derivation is different, in that we deduce the master
equation from the exact solution.

We shall discuss the question of the positivity, using the master
equation.
This argument is based on that evolution of a pure quantum state necessarily
requires \cite{ambegaokar}
\begin{equation}
\left( \,\frac{d}{dt}\,{\rm tr}\;\rho ^{2}\,\right)_{t=0} \leq  0 \,.
\label{positivity constraint} 
\end{equation}
If this inequality is violated, 
the initial pure state with ${\rm tr} \;\rho ^{2} = 1$
would evolve to a state with ${\rm tr} \;\rho ^{2} > 1$.
But with ${\rm tr}\;\rho = 1$, this  means that the state at 
$t\sim 0$ has
a negative diagonal element of the density matrix.
Using the general equation for
\( \:
\frac{d}{dt}{\rm tr}\;\rho ^{2}
\: \)
derived from the master equation, and the initial behavior,
\begin{eqnarray}
C(t) &\sim & -\frac{t^{3}}{3}\left( \,g^{(3)}(0)^{2} - g^{(5)}(0)\,\right)
\,, 
\\
D_{pp}(t) &\sim &
-\frac{\dot{U}}{4} \sim - \frac{t}{8}{\cal I}[1] < 0 \,, 
\\
D_{xp}(t) &\sim &
U \sim \frac{t^{2}}{4}{\cal I}[1] \,, 
\end{eqnarray}
one concludes that
\begin{equation}
\left( \,\frac{d}{dt}{\rm tr}\;\rho ^{2}\,\right)_{t=0} \sim 
- \, {\cal I}[1]\,\overline{(\Delta q)^{2}}\,t < 0
\,.
\end{equation}
We introduced the following notation:
\begin{equation}
{\cal I}[\,f(\omega )\,] \equiv 
\int_{\omega _{c}}^{\infty }d\omega \,
\coth (\frac{\beta \omega }{2})r(\omega )f(\omega ) \,.
\end{equation}
In this sense the correct approach satisfies the positivity constraint
at this level.
A more direct and general proof of the positivity will be given elsewhere
\cite{jmy 97-1}.

Comparison to the pole approximation, or the local friction approximation,
at $t \sim 0$
would be instructive. Here as the initial behavior one has
\begin{eqnarray}
C(t) = 2\gamma \,, \hspace{0.5cm} 
D_{pp}(t) = O[t]\,, \hspace{0.5cm} 
D_{xp}(t) = O[t^{2}] \,.
\end{eqnarray}
Hence at $t \sim 0$
\begin{eqnarray}
\frac{d}{dt}\,{\rm tr}\;\rho ^{2} &\sim & 2\gamma \,{\rm tr}\;\rho ^{2}
\,.
\end{eqnarray}
This clearly violates the positivity constraint (\ref{positivity constraint}).

For further discussion we take as the initial state 
a product of thermal states, a system of
temperature $T_{0} = 1/\beta _{0}$ 
and an environment of temperature $T = 1/\beta $.
We may take $T_{0} = T$ when we apply to the decay process of
excited level initially in thermal equilibrium.
On the other hand, 
in the limit of $T_{0} \rightarrow 0$ it describes the ground state
of the system harmonic oscillator.

After a series of straightforward
Gaussian integration we find the reduced density matrix as a function
of $X$ and $\xi $, of the form,
\begin{eqnarray}
&&
\rho ^{(R)}(X_{f}\,, \xi _{f}) = 2\sqrt{\frac{{\cal A}}{\pi }}
\exp [\,-{\cal A}X_{f}^{2} - {\cal B}\xi _{f}^{2} +
i{\cal C}\,X_{f}\xi _{f}\,] \,,
\label{density matrix} 
\\
&&
{\cal A} = \frac{1}{8 I_{1}}\,, \hspace{0.5cm} 
{\cal B} = \frac{1}{2}\, (\,I_{3} - \frac{I_{2}^{2}}{I_{1}}\,) \,, 
\hspace{0.5cm} 
{\cal C} = \frac{I_{2}}{2I_{1}} \,, 
\\
&&
I_{1} =
{\cal I}[\,|h(\omega \,, t)|^{2}\,] + 
\frac{1}{2\omega _{0}}\coth (\frac{\beta _{0}\omega _{0}}{2})\,
(\dot{g}^{2} + \omega _{0}^{2}g^{2})
\,, 
\\
&&
I_{2} =
\Re {\cal I}[\,h(\omega \,, t)k^{*}(\omega \,, t)\,] +
\frac{1}{2\omega _{0}}\coth (\frac{\beta _{0}\omega _{0}}{2})\,
\dot{g}\,(\stackrel{..}{g} + \omega _{0}^{2}g)
\,, 
\\
&&
I_{3} =
{\cal I}[\,|k(\omega \,, t)|^{2}\,]
+ \frac{1}{2\omega _{0}}\coth (\frac{\beta _{0}\omega _{0}}{2})\,
(\stackrel{..}{g}^{2} + \omega _{0}^{2}\dot{g}^{2}) \,.
\end{eqnarray}
$\omega _{0}$ is a reference frequency taken as that of the initial system
state, and equated here to the initial bare frequency. 
If one so desires, either the renormalized 
$\omega _{R}$ or the pole $\bar{\omega }$ may be taken as another
choice. But we imagine the situation a small system was added to a large
environment at some time, 
its mutual interaction being absent prior to the initial time.
In this circumstance it is appropriate to take $\omega _{0}$ as the reference
frequency.
Since dependence on the initial state dies away quickly as time passes,
the choice of the initial reference is not crucial for determining
the behavior of states at late times.
Both of $h(\omega \,, t)$ and $k(\omega \,, t)$ are already
defined in the preceeding subsection.
The density matrix $\rho ^{(R)}$ from which any physical quantity at
time $t$
can be computed has explicitly been given by the discontinuity,
$H(\omega )$ or $r(\omega )$.

The diagonal part of the reduced density matrix with $q = q'$, namely
with $\xi = 0$, is
\begin{eqnarray}
\rho ^{(R)}(X_{f} = 2q \,, 0) = 2\,\sqrt{\frac{{\cal A}}{\pi }}\,
\exp [\,-\,4{\cal A}\,q^{2} \,] \,.
\end{eqnarray}
The width of the Gaussian peak of $\approx 1/\sqrt{{\cal A}}$ is a measure of
how the system behaves. For instance, if the width increases with time,
the system is more excited than originally prepared, while if it
decreases, it is more deexcited.

The basic quantities that appear in the reduced density matrix are
related to expectation values of the coordinate and the momentum
operators at the same moment by
\begin{eqnarray}
\langle\, q^{2}\, \rangle &=& \frac{1}{8{\cal A}} = I_{1} \,, 
\\ 
\langle\, p^{2}\, \rangle &=&
2{\cal B} + \frac{{\cal C}^{2}}{2{\cal A}} = I_{3} \,, 
\\ 
\langle\, \frac{1}{2}\, (\,qp + pq \,)\, \rangle &=&
\frac{{\cal C}}{4{\cal A}} = I_{2} \,. 
\end{eqnarray}
Thus one may write the density matrix as
\begin{eqnarray}
&&
\rho ^{(R)}(X_{f}\,, \xi _{f}) =
\nonumber 
\\ && \hspace*{-1cm}
 \sqrt\frac{1}{2\pi\, \langle q^{2} \rangle}
\,\exp [\,- \,\frac{1}{8\langle q^{2} \rangle}\,X_{f}^{2} - 
\left( \,\frac{\langle p^{2} \rangle}{2} - 
\frac{\langle qp + pq \rangle^{2}}{8 \langle q^{2} \rangle}\,\right)
\,\xi _{f}^{2} + i\frac{\langle qp + pq \rangle}{4\langle q^{2} \rangle}
\,X_{f}\xi _{f}\,] \,.
\end{eqnarray}
The reduced density matrix is thus characterized by expectation
values of quadratic operators, just as in the case of pure Gaussian
system without the environmental effect.

It is sometimes useful to transform the density matrix in the
configuration space to the Wigner function $f_{W}(x\,, p)$,
\begin{eqnarray}
f_{W}(x \,, p) & \equiv& \int_{-\infty }^{\infty }\,d\xi \,
\rho^{(R)}(2x \,, \xi )\,e^{-\,i\,p \xi }
 \,, \\
&=&
\sqrt{\frac{4{\cal A}}{{\cal B}}}\,\exp [\,-\,4{\cal A}\,x^{2}
- \frac{(p - 2{\cal C}\,x)^{2}}{4{\cal B}}\,] \,.
\end{eqnarray}
The Wigner function is expected to give the probability distribution
in the phase space $(x\,, p)$ when the semi-classical picture is valid.
Expectation value of the number operator, namely the occupation number, 
in terms of the reference frequency,
equated to the pole location $\bar{\omega} $ here,
is calculated most easily from the Wigner function:
\begin{eqnarray}
\langle n \rangle &\equiv&  
\langle \,-\,\frac{1}{2\bar{\omega}}\,\frac{d^{2}}{dq^{2}} +
\frac{\bar{\omega}}{2}\,q^{2} - \frac{1}{2}\, \rangle 
= \frac{{\cal B}}{\bar{\omega}}
 + \frac{{\cal C}^{2}}{4\bar{\omega}\,{\cal A}}
 + \frac{\bar{\omega}}{16{\cal A}} - \frac{1}{2} 
\nonumber \\ 
&=&
\frac{1}{2\bar{\omega }}\,(\,I_{3} + \bar{\omega }^{2}
I_{1}\,) - \frac{1}{2}
\,. \label{number operator} 
\end{eqnarray}
It consists, except the trivial $\frac{1}{2}$, 
of two terms, the term $\bar{\omega}/(16\,{\cal A})$ from the Gaussian
width of the diagonal density matrix element and the rest from
the kinetic term $-\,\frac{d^{2}}{dq^{2}}$.
This formula of course agrees with that of the previous derivation in
the operator method.
We shall discuss the rate of particle production in more detail
when we turn to the specific application, namely the case 
of periodic frequency $\omega ^{2}(\tau )$.

An important measure to explore the behavior of the quantum state
under the environment action is the effective entropy of the subsystem
under study. 
Even if the entire system of the subsystem plus the environment
is in a pure quantum state, the subsystem can have a nonvanishing entropy,
since one cannot measure the environment and traces out its freedom.
The reduced density matrix derived in the path integral approach is
especially suited to calculation of the subsystem entropy \cite{hjmy 96}.
We define as usual the entropy of the system as
(henceforth we omit the index $R$ for the reduced density matrix)
\begin{equation}
S = -\,{\rm tr}\;\rho \ln \rho \,, 
\end{equation}
where the trace operation is performed on the system variable $q$.
A useful device to compute the logarithmic matrix is to convert it to
a power series according to the formula like
\begin{equation}
S = -\,{\rm tr}\;
\left( \,\frac{d}{ds}\,\frac{1}{\Gamma (3-s)}\,\rho ^{3}\,
\int_{0}^{\infty }\,du\,u^{2-s}\,e^{-\rho u}\,\right)_{s = 1} \,.
\end{equation}
The arbitrary power 
\( \:
{\rm tr}\;\rho ^{n}
\: \)
is calculable as
\begin{eqnarray}
{\rm tr}\;\rho ^{n} &=& \int\,\left( \prod_{i}^{n}\,dq_{i} \right)\,
\rho (q_{1}+q_{2}\,, q_{1}-q_{2})\cdots \rho (q_{n}+q_{1}\,, 
q_{n}-q_{1}) \nonumber \\
&=&
\frac{(4{\cal A})^{n/2}}{(\,\sqrt{{\cal B}} + \sqrt{{\cal A}}\,)^{n} 
- (\,\sqrt{{\cal B}} - \sqrt{{\cal A}}\,)^{n}} \,.
\end{eqnarray}

A more efficient way to compute the entropy is first to observe,
with the presence of the power series expansion, independence
of the factor ${\cal C}$ in the entropy formula. This justifies neglect of
the ${\cal C}$ term (by setting ${\cal C} = 0$)
in the density matrix (\ref{density matrix}) for
computation of the entropy, and only for this purpose.
The next step is to identify this density matrix with ${\cal C} = 0$
as an equivalent harmonic oscillator system of frequency $\tilde{\omega }$
under a thermal bath of temperature $1/\tilde{\beta }$:
\begin{equation}
\sqrt{\,\frac{\tilde{\omega}}{\pi \,\coth 
(\tilde{\beta}\tilde{\omega} /2)}\,}\,
\exp \left( \,
- \,\frac{\tilde{\omega}}{4}\,
\coth (\frac{\tilde{\beta}\tilde{\omega}}{2})\,
\xi _{f}^{2} - \frac{\tilde{\omega}}{4}\,
\tanh  (\frac{\tilde{\beta}\tilde{\omega}}{2})
\,X_{f}^{2}
\,\right)  \,.
\end{equation}
The equivalence is possible only for 
\( \:
{\cal B} > {\cal A}
\: \)
and is established by the parameter relation,
\begin{eqnarray}
\tilde{\omega } = 4\,\sqrt{{\cal A}{\cal B}} \,, \hspace{0.5cm} 
\tanh \frac{\tilde{\beta }\tilde{\omega }}{2} = 
\sqrt{\frac{{\cal A}}{{\cal B}}} \,.
\end{eqnarray}
With this identification the equivalent temperature and the entropy are
given by \cite{hjmy 96} 
\begin{eqnarray}
\tilde{T} &=& \frac{4\,\sqrt{{\cal A}{\cal B}}}{\ln \left( \,
(\sqrt{{\cal B}} + \sqrt{{\cal A}})/
(\sqrt{{\cal B}} - \sqrt{{\cal A}}) \,\right)} \,, \\
S &=& \frac{\tilde{\beta }\tilde{\omega }}{e^{\tilde{\beta }\tilde{\omega }}
- 1} - \ln (\,1 - e^{-\,\tilde{\beta }\tilde{\omega }}\,) \nonumber \\
&=& \frac{1}{2}\, (x - 1)\ln \frac{x + 1}{x - 1} + \ln \frac{x + 1}{2}
\,, 
\label{entropy formula} 
\end{eqnarray}
with
\begin{equation}
x = \sqrt{\frac{{\cal B}}{{\cal A}}} = 2\,\sqrt{\,I_{1}I_{3} - I_{2}^{2}\,}
=
\sqrt{\,4\,\langle q^{2} \rangle\langle p^{2} \rangle
- \langle qp + pq \rangle^{2} \,} 
\,.
\end{equation}
The entropy $S$ is a monotonic function of the single variable
\( \:
x = \sqrt{{\cal B}/{\cal A}} \,.
\: \)
Its limiting values are
\begin{eqnarray}
S &\rightarrow & -\,\frac{1}{2}\, (x - 1)\,\ln \frac{x - 1}{2} \,, 
\hspace{0.5cm} {\rm as} \; x \rightarrow 1^{+} \,, 
\\
&\rightarrow & \ln \frac{x}{2} \,, \hspace{0.5cm} 
{\rm as} \; x \rightarrow \infty \,.
\end{eqnarray}

It is interesting to note that the entropy initially increases
irrespective of the relation between the two temperatures of the
subsystem and the environment, $T_{0}$ and $T$.
This can be proved by first writing down the time evolution equation
for the quantity
\( \:
x^{2} = 4\,\langle q^{2} \rangle\langle p^{2} \rangle
- \langle qp + pq \rangle^{2} \, :
\: \)
\begin{equation}
\frac{dx^{2}}{dt} = -\,2C(t)\,x^{2} - 16 D_{pp}(t)\,\langle \,q^{2}\, \rangle
- 8 D_{xp}(t)\,\langle \,qp + pq \, \rangle \,.
\end{equation}
As $t \rightarrow 0^{+}$, the dominant term in the right hand side
is the $D_{pp}(t)$ term. Thus at $t \sim 0$
\begin{eqnarray}
\frac{dx^{2}}{dt} \sim 2\,{\cal I}[1]\,t\,\langle \,q^{2}\, \rangle
> 0 \,.
\end{eqnarray}
This proves the entropy increase.

When 
\( \:
{\cal A} \ll {\cal B} 
\: \) 
holds, considerable simplification follows:
\begin{eqnarray}
\tilde{\omega } &=& 4\sqrt{{\cal A}{\cal B}} \,, \\
\tilde{T} 
&\sim &
2\,{\cal B} \,, \\
S 
&\sim &
\frac{1}{2}\, \ln \frac{{\cal B}}{4{\cal A}} + 1 \,.
\end{eqnarray}

When ${\cal B} = {\cal A}$, or equivalently
\( \:
\langle p^{2} \rangle \langle q^{2} \rangle =
\langle qp^{2}q \rangle \,, 
\: \)
the density matrix is factorizable into a product of 
\( \:
\psi (q)\,\psi ^{*}(q')
\: \),
since ${\rm tr}\;\rho ^{n} = 1$.
This is a pure quantum state, having the entropy
\( \:
S = 0 \,.
\: \)

When the initial subsystem temperature $T_{0} = 1/\beta _{0}$ 
is nonvanishing, the entropy is
\begin{equation}
S_{i} = \frac{\beta _{0}\omega _{0}}{e^{\beta _{0}\omega _{0}} - 1}
- \ln (\,1 - e^{-\beta _{0}\omega _{0}}\,) \,,
\end{equation}
initially. For $T_{0} \gg \omega _{0}$,
\( \:
S_{i} \approx \ln \frac{T_{0}}{\omega _{0}} \,.
\: \)
It ends finally with
\begin{equation}
S_{f} = \ln \frac{\tilde{T}}{\tilde{\omega }} \,,
\end{equation}
for ${\cal B} \gg {\cal A}$.
As will be made clear shortly, this is equal to 
\( \:
\ln \frac{T}{\bar{\omega }}
\: \)
in the high temperature limit.
This variation, from $\ln (T_{0}/\omega _{0})$ to $\ln (T/\bar{\omega })$,
is a result one would naively expect.

Existence of the equivalent harmonic oscillator basis for 
${\cal B} \geq {\cal A}$ actually implies much more than a convenient
means of the entropy computation.
We have found the basis of the states that diagonalize
the density matrix at any instant of time: 
it is the Fock basis $|\,n \,\rangle $ of
the equivalent harmonic oscillator of frequency $\tilde{\omega }$
with diagonal elements:
\begin{equation}
\rho _{n} = 2\sinh (\frac{\tilde{\beta }\tilde{\omega }}{2})\,
e^{-\,\tilde{\beta }\tilde{\omega }\,(n + \frac{1}{2})} =
\frac{\sqrt{{\cal B} }- \sqrt{{\cal A}}}{2\sqrt{{\cal A}}}\,
\left( \,\frac{\sqrt{{\cal B} }- \sqrt{{\cal A}}}
{\sqrt{{\cal B} }+ \sqrt{{\cal A}}}\,\right)^{n}\,.
\end{equation}
Since the parameters, ${\cal A}$ and ${\cal B}$, are time dependent,
the diagonalization of the density matrix is instantaneous.
One thus thinks of the subsystem temperature varying with time as
\begin{equation}
\tilde{T} = \frac{1}{2\,\langle q^{2}(t) \rangle}\,\frac{x(t)}
{\ln \frac{x(t) + 1}{x(t) - 1}} \,, 
\end{equation}
however with frequency also changing as
\begin{equation}
\tilde{\omega } = \frac{x(t)}{2\langle q^{2}(t) \rangle} \,.
\end{equation}

To explore a more detailed nature of the final state, 
let us work out the frequency distribution of populated harmonic oscillator
levels. Suppose that one asks the probability of finding the system in
the unit energy interval around
\( \:
\omega = \frac{p^{2}}{2} + \frac{\omega _{0}^{2}}{2}\,q^{2} \,.
\: \)
We get the energy distribution function $f(\omega )$ 
\cite{hjmy 96} from the Wigner function:
\begin{eqnarray}
f(\omega ) &\equiv& \int\,\frac{dq\,dp}{2\pi }\,f_{W}(q\,, p)\,
\delta (\,\omega - \frac{p^{2}}{2} - \frac{\omega _{0}^{2}}{2}\,q^{2}\,)
\\
&=&
\frac{2}{\omega _{0}}\,\sqrt{\frac{{\cal A}}{{\cal B}}}\,
\exp \left( \,-\,\frac{\omega}{\omega _{0}}\, 
(\,\frac{4{\cal A}}{\omega _{0}} + \frac{4{\cal C}^{2} + \omega _{0}^{2}}
{4\omega _{0}\,{\cal B}}\,)\,\right)\,I_{0}(\,\frac{\omega }{\omega _{0}}
\,z\,) \,, \label{energy distribution} \\
z &=& \left( \,(\,\frac{4{\cal A}}{\omega _{0}} 
+ \frac{4{\cal C}^{2} + \omega _{0}^{2}}
{4\omega _{0}\,{\cal B}}\,)^{2} - \frac{4{\cal A}}{{\cal B}}\,\right)^{1/2}
\,.
\end{eqnarray}
Here 
\begin{equation}
I_{0}(w) = \frac{1}{\pi }\,\int_{0}^{\pi }\,d\theta \,e^{-\,w\cos \theta }
\end{equation}
is the modified Bessel function, that behaves as
\begin{equation}
I_{0}(w) \:\rightarrow  \: \frac{1}{\sqrt{\,2\pi w\,}}\,e^{w} 
\hspace{0.3cm} ({\rm as \;} w \rightarrow \infty ) \,, \hspace{0.5cm} 
\:\rightarrow  \: 1 \hspace{0.3cm} ({\rm as \;} w \rightarrow 0) \,.
\end{equation}
$\omega _{0}$ is again the reference frequency not to be confused
with the original subsystem frequency.

This distribution differs from the thermal one and has some
interesting features.
At $\omega \rightarrow \infty $, it decreases as
\begin{eqnarray}
f(\omega ) &\rightarrow& \frac{2}{\sqrt{\,2\pi \,\omega _{0}\omega \,z\,}}
\,\sqrt{\,\frac{{\cal A}}{{\cal B}}\,}\,e^{-\,\omega \,D}\,, 
\\
D &=& \frac{1}{\omega _{0}}\,
\left[\,
\frac{4{\cal A}}{\omega _{0}} + \frac{4{\cal C}^{2} + 
\omega _{0}^{2}}{4\omega _{0}\,{\cal B}} 
- \left( \,(\,\frac{4{\cal A}}{\omega _{0}} +
\frac{4{\cal C}^{2} + \omega _{0}^{2}}{4\omega _{0}\,{\cal B}}\,)^{2} 
- \frac{4{\cal A}}{{\cal B}} \,\right)^{1/2}  \,\right]
\,,
\end{eqnarray}
assuming $z \neq 0$.
The high frequency tail of this distribution decreases more rapidly
than that of the Boltzmann distribution, $e^{-\,\beta \omega }$.
On the other hand, at $\omega = 0$
\begin{equation}
f(0) = \frac{2}{\omega _{0}}\,\sqrt{\frac{{\cal A}}{{\cal B}}} \,.
\end{equation}
The average energy of the distribution is computed as
\begin{eqnarray}
\langle \omega  \rangle &\equiv &
\int_{0}^{\infty }\,d\omega \,\omega \,f(\omega )
\nonumber \\
&=&
\frac{\omega _{0}{\cal B}}{4{\cal A}}\,\left( \,
\frac{4{\cal A}}{\omega _{0}} + \frac{4{\cal C}^{2} + 
\omega _{0}^{2}}{4\omega _{0}\,{\cal B}}
\,\right) \,. \label{average energy} 
\end{eqnarray}
This average satisfies the relation,
\begin{equation}
\langle \omega  \rangle = \omega _{0}\,(\,\langle n \rangle + \frac{1}{2}\,)
\,,
\end{equation}
when $\bar{\omega }$ is replaced by $\omega _{0}$ 
in eq.(\ref{number operator}).

The exceptional case of $z = 0$ with an exact exponential form
is actually important, since it is realized in thermal equilibrium. 
This is possible only when
\( \:
{\cal C} = 0 \,, \hspace{0.3cm} {\rm and} \hspace{0.3cm}
\sqrt{\,{\cal A}{\cal B}\,} = \omega _{0}/4 \,,
\: \)
giving
\begin{eqnarray}
f(\omega ) &=& \frac{e^{-\,\omega /T}}{T} 
 \,, \\
\langle \omega  \rangle &=& T \,.
\end{eqnarray}

The normalized spectrum shape written in terms of a scaled energy,
$x = \omega /\langle \omega  \rangle$, is characterized by a single parameter
$\delta $,
\begin{eqnarray}
f_{0}(x\,; \delta ) &=& \frac{1}{\delta }\,e^{-\,x/\delta ^{2}}\,
I_{0}(\frac{\sqrt{1 - \delta ^{2}}}{\delta ^{2}}\,x) \\
&=&
\frac{1}{\pi \delta }\,e^{-\,x/\delta ^{2}}\,
\int_{0}^{\pi }\,d\theta \,\exp [\,-\,x\,\frac{\sqrt{1 - \delta ^{2}}}
{\delta ^{2}}\,\cos \theta \,] \,, 
\label{universal spectrum} 
\end{eqnarray}
with which the energy distribution is given by
\begin{eqnarray}
&&
f(\omega ) = \frac{1}{\langle \omega  \rangle}\,
f_{0}\left(\,\frac{\omega }{\langle \omega  \rangle} \,; \delta\,\right ) 
\,, \\
&&
\hspace*{-2cm}
\delta = 
2\sqrt{\,\frac{{\cal A}}{{\cal B}}\,}\,
\left(\,\frac{4{\cal A}}{\omega _{0}} + \frac{4{\cal C}^{2} + 
\omega _{0}^{2}}{4\omega _{0}\,{\cal B}} \,\right)^{-\,1} 
=
\frac{\omega _{0}\,\sqrt{\,4\,\langle q^{2} \rangle\langle p^{2} \rangle
- \langle qp + pq \rangle^{2} \,}}{\langle p^{2} \rangle + \omega _{0}^{2}
\,\langle q^{2} \rangle} = \frac{x}{2\langle n \rangle + 1}
\,.
\end{eqnarray}
This function has the following limiting behavior:
\begin{eqnarray}
f_{0}(x\,; \delta ) &\rightarrow  & e^{-\,x} \; {\rm as} \;
\delta \:\rightarrow\: 1 \,, \\
f_{0}(x\,; \delta ) &\rightarrow & \frac{1}{\sqrt{2\pi \,x}}\,e^{-\,x/2}
\; {\rm as} \; \delta \:\rightarrow  \: 0 \,.
\end{eqnarray}
The first moment $\overline{x}$ happens to be the same in the
two limits of $\delta = 1 $ and $\delta = 0$.
The main difference of these two limiting distributions is in the low
$x$ part.
We call $\delta $ the spectral shape parameter.

The asymptotic late time behavior of the reduced density matrix
is determined by 
\( \:
h(\omega \,, \infty ) = \int_{0}^{\infty }\,
d\tau \,g(\tau )e^{-\,i\omega \tau } \,, 
\: \)
which is equal to the boundary value,
\( \:
F(\omega - i0^{+} ) \,,
\: \)
thus giving the already known relations,
\( \:
r(\omega )|h(\omega \,, \infty )|^{2} = H(\omega ) \,, 
\hspace{0.5cm} 
r(\omega )|k(\omega \,, \infty )|^{2} = \omega ^{2}H(\omega )
\,.
\: \)
Thus, at asymptotic late times
\begin{eqnarray}
{\cal A} &\rightarrow  &
\frac{1}{8}\,
\left( \,
\int_{\omega _{c}}^{\infty }
d\omega \,\coth (\frac{\beta \omega }{2})\,H(\omega )\,\right)^{-1}
\,, \label{asymptotic a} 
\\
{\cal B} &\rightarrow &
\frac{1}{2}\,
\int_{\omega _{c}}^{\infty }d\omega \,\coth (\frac{\beta \omega }{2})\,
\omega ^{2}H(\omega )
\,, \label{asymptotic b} 
\\ 
{\cal C} &\rightarrow&    0 \,.
\end{eqnarray}
Equivalently,
\begin{eqnarray}
\langle q^{2} \rangle &\rightarrow & 
\int_{\omega _{c}}^{\infty }
d\omega \,\coth (\frac{\beta \omega }{2})\,H(\omega )
\\
\langle p^{2} \rangle &\rightarrow & 
\int_{\omega _{c}}^{\infty }d\omega \,\coth (\frac{\beta \omega }{2})\,
\omega ^{2}H(\omega )
\,.
\end{eqnarray}
These agree with those in the operator approach.
In these computations the analytic structure and only that is
important. 

In the high temperature limit 
\begin{equation}
\langle q^{2} \rangle \sim \frac{T}{\omega _{R}^{2}} \,, \hspace{0.5cm} 
\langle p^{2} \rangle \sim T \,.
\end{equation}
The renormalized frequency $\omega _{R}$ is defined in the preceeding
subsection, eq.(\ref{renormalized frequency}) and is given by
\begin{equation}
\omega _{R}^{2} = \left( \, \int_{\omega _{c}}^{\infty }\,d\omega 
\,\frac{2H(\omega )}{\omega } \,\right)^{-1} \,.
\end{equation}
In this contribution the continuous integral during the period of 
the power law decay is numerically subdominant, suppressed by the factor
$1/T^{2}$ relative to the one from the exponential period.
On the other hand, at low temperatures the contribution from
the threshold region, $\omega \approx \omega _{c}$, cannot be ignored,
giving the dominant contribution to the power law period, as noted
in the preceeding subsection.

It may be useful to write  crude formulas for asymptotic values
in the Breit-Wigner approximation,
using
\( \:
2\omega \,H(\omega ) \approx \delta (\omega - \bar{\omega }) \, :
\: \)
\begin{eqnarray}
&& \hspace*{-1cm}
\langle \,q^{2}(\infty )\, \rangle \approx \frac{1}{2\bar{\omega }}\,
\coth (\frac{\beta \bar{\omega }}{2}) \,, \hspace{0.5cm} 
\langle \,p^{2}(\infty )\, \rangle \approx \frac{\bar{\omega }}{2}\,
\coth (\frac{\beta \bar{\omega }}{2}) \,, \hspace{0.5cm} 
\tilde{\omega }(\infty ) \approx \bar{\omega } \,, 
\\ && \hspace*{-1cm}
\tilde{T}(\infty ) \approx T \,, \hspace{0.5cm} 
x(\infty ) \approx \coth (\frac{\beta \bar{\omega }}{2}) \,, \hspace{0.5cm} 
\delta(\infty ) \approx \frac{2\bar{\omega }\omega _{0}}{\bar{\omega }^{2} +
\omega _{0}^{2} } \,.
\end{eqnarray}
These are not precise, but they are nevertheless instructive
to understand precise formulas for these quantities.
$\delta (\infty ) < 1$ for $\omega _{0} \neq \bar{\omega }$.
But if one takes the reference frequency equal to $\bar{\omega }$,
one gets
\( \:
\delta (\infty ) = 1 \,, 
\: \)
in complete agreement with thermal equilibrium.

\vspace{0.5cm} 
In closing this section,
we would like to stress that the two approaches we described,
the operator and the path integral, are both useful and complement
each other.
In both approaches the maximal use of the extended analyticity
has been instrumental to completely clarify the details of
the linear open system.
The model of the linear open system is thus analytically solvable.

\vspace{1cm} 
\begin{center}
\begin{Large}
{\bf \lromn 3 Particle Decay in Thermal Medium}
\end{Large}
\end{center}

Let us apply these considerations to the decay of unstable particle;
\( \:
\varphi \rightarrow \chi + \chi \,.
\: \)
We assume that the decay product $\chi $ is a part of thermal components that
make up the environment.
The parent particle $\varphi $ of mass $M$
may or may not be in thermal equilibrium
with the rest of medium: since we focus on the late time behavior,
the initial state dependence may be ignored.

Let me first explain how the field theory may be incorporated
in the framework so far discussed.
The point is that the infinitely many environment variable $Q_{a}$ 
can actually be a composite field
of more fundamental fields. 
We consider as the system variable $q_{k}$
a Fourier component of some scalar field $\varphi (x)$ in
a fully relativistic field theory.
As a simplest field theory model that may describe the decay process,
take a Yukawa type of interaction with a new bose field $\chi (x)$;
\begin{equation}
L_{{\rm int}} = \int_{V}\,d^{3}x\,{\cal L}_{{\rm int}} \,, \hspace{0.5cm} 
{\cal L}_{{\rm int}} = -\,\frac{\mu }{2}\,\varphi \chi ^{2}
\,,
\end{equation}
with $\mu $ a parameter of mass dimension. 
$V$ is the normalization volume.
By identifying the Fourier-mode with
\begin{equation}
q_{k}\,e^{i\vec{k}\cdot \vec{x}} = 
\frac{1}{\sqrt{2\omega _{k}\,V}\,} \;(\,a_{k} + a^{\dag }_{-k}\,)\,
e^{i\vec{k}\cdot \vec{x}} \,, 
\end{equation}
we deduce the environment variable $Q_{k}(\omega )$ coupled to it,
\begin{equation}
\int\,d\omega \,c(\omega )Q_{k}(\omega ) =
\frac{\mu }{2}\,\int_{V}\,d^{3}x\,\chi ^{2}(x)\,e^{-\,i\vec{k}\cdot \vec{x}}
\,. 
\end{equation}
The continuous label $\omega $ of the environment variable $Q_{k}(\omega )$
is thus identified to 
the internal configuration of two body states with a given total momentum
$\vec{k}$.
This field theory model was introduced in ref \cite{hjmy 96}.
Henceforth we regard the field $\chi $ as a fundamental variable
of the environment, assumed to be in thermal equilibrium.

The correlation function $\alpha_{k}(\tau )$, or more conveniently its
Fourier transform $\alpha_{k}(\omega )$, can be computed as in the 
real-time formalism at finite temperatures. To lowest order in $\mu ^{2}$
\begin{equation}
i\,\alpha_{k} (\tau - s)  = i\,(\frac{\mu }{2})^{2}\,
\int_{V}\,d^{3}x\,e^{i\vec{k}\cdot \vec{x}}\;
{\rm tr}\;
\left( \,T\,[\,\chi ^{2}(\vec{x}\,, \tau )\,
\chi ^{2}(\vec{0}\,, s)\,]\,\rho_{\beta } 
\,\right) 
\end{equation}
is calculable in terms of the $\chi-$propagator in the momentum space,
\begin{equation}
i\,\left( \,\frac{1}{\omega ^{2} - \vec{k}^{2} - m_{\chi }^{2} +
i\epsilon } - \frac{2\pi i}{e^{\beta \omega _{k}} - 1}\,
\delta (\omega ^{2} - \vec{k}^{2} - m_{\chi }^{2} )\,\right) \,.
\end{equation}
We omit the suffix $\chi $ for the $\chi-$mass $m_{\chi }$ in what
follows.

As in the $T=0$ field theory it is easiest to first compute
the discontinuity instead of the full diagramatic contribution.
Physically the most transparent way 
to compute the imaginary part of the self-energy diagram,
\( \:
\Im \,\Pi (\omega ) \,, 
\: \)
is to use the analytically continued expression  \cite{weldon}
from the imaginary-time formalism.
The analytically continued imaginary part, or more precisely the
discontinuity, directly gives the response weight via
\begin{equation}
r (\omega ) = 2\Im \,\Pi (\omega ) = 2\omega\, \Gamma (\omega )
\,, \hspace{0.5cm} 
\Im \,\Pi (\omega ) \equiv  \frac{1}{2i}\,\left( \,
\Pi (\omega - i\epsilon ) - \Pi (\omega + i\epsilon )\,\right) \,,
\end{equation}
where $\Gamma (\omega )$ is interpreted as a decay rate in
thermal medium.

The result for the $\chi $ loop diagram at finite temperatures is well known 
\cite{weldon}, \cite{hjmy 96},
and physically consists of two parts; the process $\varphi \leftrightarrow 
\chi + \chi $ in the region of
$\omega > \sqrt{k^{2} + 4m^{2}}$ and the other process
\( \:
\varphi + \chi \leftrightarrow \chi 
\: \)
(forbidden when all particles are on the mass shell, but allowed
in thermal environment)
in $0 < \omega < k$, where $m$ is the daughter $\chi $ mass.
The response weight $r(\omega )$ does not vanish for
\( \:
|\omega | > \sqrt{\vec{k}^{2} + 4m^{2}}
\: \)
and
\( \:
|\omega | < |\vec{k}|
\: \)
from the kinematics of the decay and the inverse decay of particles 
off the mass shell, with the constraint of the momentum conservation.
Thus a gap of the spectrum exists in
\( \:
k < |\omega | < \sqrt{\vec{k}^{2} + 4 m^{2}}
\,. 
\: \)
The finite non-vanishing mass of the daughter particle ($m \neq 0$)
is important for the existence of the gap and for associated
physical consequences that follow.

Let us explain some details of the calculation of the response
weight in the subthreshold region of 
\( \:
|\omega | < k \,.
\: \)
One loop contribution from the imaginary-time formalism is given by
\cite{weldon}
\begin{eqnarray}
\Im \Pi\, (\omega ) &=& \frac{\mu ^{2}}{16\pi k}\,
\int_{- \omega _{-}}^{\infty }\,dE\,\left( \,n(E) - n(E + \omega )\,\right)
\,, \label{subthreshold imaginary part} \\
\omega _{\pm } &=&
\frac{\omega }{2} \pm \frac{k}{2}\,
\sqrt{\,1 - \frac{4m^{2}}{\omega ^{2}-k^{2}}\,} \,,
\end{eqnarray}
where
\begin{equation}
n(E) = \frac{1}{e^{\beta E} - 1}
\end{equation}
is the Planck distribution function of $T = 1/\beta $.
Since
\begin{equation}
n(E) - n(E + \omega ) = n(E)(\,1 + n(E + \omega )\,) -
n(E + \omega )(\,1 + n(E)\,) \,, 
\end{equation}
the imaginary part (\ref{subthreshold imaginary part}) for
$|\omega | < k$ is a sum of the two contributions,
\( \:
\chi + \varphi \rightarrow \chi 
\: \)
and its inverse process that is allowed to occur in thermal medium.
Note that $\varphi $ can be off the mass shell: 
\( \:
\omega ^{2} - \vec{k}^{2} \neq 
\: \)
the $\varphi $ mass$^{2}$.
The factor $1 + n$ represents the effect of stimulated boson emission.

The result of computation is now summarized.
For $\omega > \sqrt{k^{2} + 4m^{2}}$ the response weight is
\cite{jmy-96-1} 
\begin{eqnarray}
r(\omega ) = \frac{\mu ^{2}}{16\pi }\,
\left( \,\sqrt{\,1 - \frac{4m^{2}}{\omega ^{2} - k^{2}}\,} +
\frac{2}{k\beta }\,\ln \frac{1 - e^{-\beta \omega _{+}}}
{1 - e^{-\beta |\omega _{-}|}}\,\right) \,. 
\label{response for decay}
\end{eqnarray}
Note that $r(\omega ) \rightarrow $ a constant
($= 2M \times $ decay rate in the rest frame of $\varphi $), 
hence the frequency
renormalization or the subtracted form of $\omega $ integral is
necessary.
For $0 < \omega < k$ only the second term in the bracket of 
Eq.(\ref{response for decay}) contributes.

A useful, and adequate approximation we exploit for subsequent estimate
is the weak coupling scheme with correct
threshold and asymptotic behaviors incorporated: 
\begin{equation}
F(z) =
\frac{1}{- z^{2} + \bar{\omega }^{2} - i\,\pi r(z)} \,.
\end{equation}
In this approximation we replaced the $\omega $ dependent 
real part of the self-energy $\Pi (\omega )$ by a constant, hence by
the constant pole location $\bar{\omega }$.

A quantity of physical interest is the fraction of remaining particles 
given by the occupation number at asymptotic late times, 
\begin{equation}
n_{k} = \frac{1}{2}\, \langle \,\frac{p_{k}^{2}}{\omega _{k}} +
\omega _{k}q_{k}^{2}\, \rangle - \frac{1}{2}
\approx  \frac{{\cal B}_{k}}{\omega _{k}} + 
\frac{\omega _{k}}{16{\cal A}_{k}} - \frac{1}{2} \,, 
\end{equation}
for each $\vec{k}$ mode ($\omega _{k} = \sqrt{\vec{k}^{2} + M^{2}}$).
The temperature dependent part of this quantity is
\begin{equation}
n^{\beta  }_{k} = 
\int_{0}^{\infty }\,d\omega \,\frac{1}{e^{\beta \omega }
- 1}\,(\omega_{k}  + 
\frac{\omega ^{2}}{\omega_{k}} )\,H(\omega \,, k) \,.
\end{equation}
One must sum over momentum $\vec{k}$ to obtain the number density of remnants.

We shall limit our discussion here to the decay that occurs 
when the parent $\varphi $ becomes non-relativistic, 
\begin{equation}
\omega_{k} \sim  M + \frac{\vec{k}^{2}}{2M} \gg T \,.
\end{equation}
This condition is relevant in interesting cosmological problems of the neutron
decay at the time of nucleosynthesis and GUT $X$ boson decay at
baryogenesis \cite{baryogenesis review96}.

Computation of the temperature dependent part of the occupation number
$n_{k}^{\beta }$ may proceed 
by deforming the contour of $\omega $ integration, in the same way
as in the discussion of $g(\tau )$ in the preceeding section.
There are then two types of contribution: one is the pole term that gives 
the usual Boltzmann suppressed contibution of $e^{-\,\beta \omega _{k}}$.
When mode-summed, it gives the number density,
\begin{equation}
\int\,\frac{d^{3}k}{(2\pi )^{3}}\,e^{-\,\beta (\,M + k^{2}/2M\,)}
= (\frac{MT}{2\pi })^{3/2}\,e^{-M/T} \,.
\end{equation}
This is the familiar Boltzmann suppressed formula.

The second one is contribution from the continuous
complex path that gives the power law behavior of temperature dependence.
A part of this contribution in the region $\omega > \sqrt{k^{2} + 4m^{2}}$ 
is analytically calculable by using
\( \:
r(\omega ) = \frac{\mu ^{2}}{16\pi } + O[m^{2}] \,, 
\: \)
valid for a small daughter mass $m$. It is
\begin{eqnarray}
n &\approx& \frac{1}{2\pi ^{2}}\,\frac{\mu ^{2}}{16\pi M^{3}}\,
\int_{0}^{\infty }\,dk\,k^{2}\,
\int_{k}^{\infty }\,d\omega \,\frac{1}{e^{\omega /T} - 1}
\nonumber 
\\ 
&=& \frac{\pi }{1440}\,\frac{\mu ^{2}T^{4}}{M^{3}} \,.
\end{eqnarray}
This calculation however ignores complicated logarithmic factors in
$r(\omega )$ of eq.(\ref{response for decay}).

We numerically computed \cite{jmy-96-1} all terms including the logarithmic
factor in $r(\omega )$ along with $O[m^{2}]$ corrections.
It turns out that the total contribution is ten times larger 
than the analytic result above: in the $m\rightarrow 0$ limit,
\begin{equation}
n \approx  2.4\times 10^{-2}\,\frac{\mu ^{2}T^{4}}{M^{3}} \,.
\end{equation}
The main part of this large contribution comes from $|\omega | < k$.
With a dimensionless constant introduced by $\mu = gM$, 
this gives, relative to the photon number density
\( \:
( = \frac{2\zeta (3)}{\pi ^{2}}\,T^{3})
\: \),
\begin{equation}
\frac{n}{T^{3}} \approx  2\times 10^{-12}\,
(\,\frac{g}{G_{F}m_{N}^{2}}\,)^{2}\,\frac{T}{M} \,.
\end{equation}
We wrote here the numerical value using $G_{F}$ 
the weak interaction constant of mass dimensions $-\,2$
(\( \:
G_{F}m_{N}^{2} \approx 10^{-\,5} 
\: \)),
as if it were relevant to the neutron decay.

One may estimate the equal time temperature $T_{{\rm eq}}$ 
at which the power contribution becomes equal
to the Boltzmann suppressed number density, to give
\begin{equation}
\frac{T_{{\rm eq}}}{M} \approx \frac{1}{33} \,, \hspace{0.5cm} 
\frac{n}{T^{3}_{{\rm eq}}} \approx 7 \times 10^{-14} \,,
\end{equation}
taking as an example 
\( \:
\mu = 10^{-5}\,M \,,
\: \)
the weal interaction strength.
This number is in an interesting range to affect nucleosynthesis, but
we should keep in mind that we did not work out the relevant three
body decay, 
\( \:
n \rightarrow p + e + \bar{\nu }_{e} \,.
\: \)

The physical interpretation of the pole term is a conventional one
in terms of the remnant created by the inverse decay $\chi + \chi 
\rightarrow \varphi $, with all relevant particles on the mass shell,
hence suppressed kinematically by the Boltzmann factor $e^{-M/T}$.
On the other hand, the contribution from the continuous contour
can only be interpreted as remnant particles far off the mass shell
that may exist in thermal equilibrium.
The usual kinetic approach such as the Boltzmann-like equation is
based on the rates computed from S-matrix elements on the mass
shell and gives the Boltzmann suppressed abundance in equilibrium
for $M \gg T$. 
Our fully quantum mechanical approach yields
a different result.

We note that the local friction approximation is equivalent to
the pole model (with identification of
\( \:
\omega _{0}^{2} + \delta \omega ^{2} = (\Re z_{0})^{2} \,, \;
\eta = -\,2\Im z_{0}
\: \))
that ignores the continuum integral around the threshold. 
Hence the pole model, or the local friction approximation, 
fails to correctly describe the off-shell remnant.

We shall mention another application of immediate interest in cosmology;
the heavy $X$ boson decay for GUT baryogenesis.
It has been argued \cite{baryogenesis review96} that there exists
a severe mass bound of order,
\begin{equation}
m_{X} > O[\alpha _{X}\,m_{{\rm pl}}] \approx 10^{16}\,{\rm GeV} \,, 
\end{equation}
to block the inverse process of the $X$ boson decay so that generation
of the baryon asymmetry proceeds with sufficient abundance of parent
$X$ particles.
The usual estimate of the mass bound mentioned above is however based
on the on-shell Boltzmann equation.
More appropriate formula in this estimate is our remnant number density,
\begin{equation}
n_{X} \approx O[2\times 10^{-2}]\,g_{X}^{2}\,\frac{T^{4}}{m_{X}} \,.
\end{equation}
(In a more realistic estimate one should consider the $X$ boson decay
into quarks and leptons. But for an order of magnitude estimate difference
in statistics is not important.)
With the GUT coupling of $g_{X}^{2}/4\pi = 1/40$, the equal temperature
is roughly
\begin{equation}
T_{{\rm eq}} \approx \frac{M}{2} \,.
\end{equation}
Thus, already at temperature of about half of the $X$ mass the Boltzmann
suppressed formula is replaced by the power formula.
The kinematical condition for baryogenesis must be reconsidered in view of our
off-shell formula.

\vspace{1cm} 
\begin{center}
\begin{Large}
{\bf \lromn 4 Parametric Resonant Particle
Production by Coherent Field Oscillation}
\end{Large}
\end{center}

We now switch to an entirely different dynamical system under dissipative
environment.
Coherent field oscillation often appears in modern cosmology.
Two main examples worthy of explicit mention 
are inflaton oscillation that gives rise to the
hot big band from the null state after inflation, and 
the moduli field oscillation
that often appears in supergravity theories.

We consiser the case in which a quantum system field
$\varphi $ is coupled to a coherent field oscillation
\( \:
\xi (t) \propto \cos (m_{\xi }t)
\: \). 
The coherent field may be viewed as an aggregate of zero-momentum
particles with some kind of precise coherence.
$m_{\xi }$ is thus the mass of these bosons.
Here this field oscillation is regarded as given, 
hence we do not discuss field damping due
to particle production of the $\varphi-$field.
Our main concern here is in the effect of thermal bath on the 
$\varphi-$particle production caused by $\xi-$oscillation.

Under this circumstance the harmonic oscillator variable is
identifed to
\( \:
q_{k} = 
\frac{1}{\sqrt{\,2\omega _{k}\,V\,}\,}\,(\,a_{k} + a^{\dag }_{-k}\,) \,, 
\: \)
where the creation and the annihilation operators  are those of
Fourier $\vec{k}-$ mode ($\propto e^{i\vec{k}\cdot \vec{x}}$) of
the system field $\varphi (x)$.
For the oscillator-system interaction we take the quartic coupling
given by
\begin{equation}
\frac{1}{2}\, g^{2}\xi ^{2}\varphi ^{2} \,,
\end{equation}
primarily because this case has been analyzed in considerable detail
without taking into account the environment effect.
The $\vec{k}-$mode variable $q_{k}$ then obeys the evolution equation,
\begin{equation}
\left( \,\frac{d^{2}}{d\tau ^{2}} + \omega_{k} ^{2}(\tau )\,\right)
\,q_{k}(\tau ) = 0 \,, \hspace{0.5cm} 
\omega_{k} ^{2}(\tau ) = \vec{k}^{2} + m_{\varphi }^{2}
+ g^{2}\xi _{0}^{2}\,\cos^{2}(m_{\xi }\tau )
\,. \label{original mahieu} 
\end{equation}
$\xi _{0}$ is the amplitude of oscillation.
The standard form of this type of equation is called the Mathieu
equation and is usually written in dimensionless units:
\begin{eqnarray}
&&
\left( \,\frac{d^{2}}{dz^{2}} + h + 2\theta \,\cos (2z)\,\right)
\,q_{k}(z) = 0 \,, \\
&&
\hspace*{0.5cm} 
z = m_{\xi }\tau \,, \hspace{0.5cm} 
h = \frac{\vec{k}^{2}+m_{\varphi }^{2}}{m_{\xi }^{2}} + 
2\theta \,, \hspace{0.5cm} 
\theta = \frac{g^{2}\xi _{0}^{2}}{4m_{\xi }^{2}} \,.
\end{eqnarray}

It is well known that this quantum system exhibits instability in infinitely
many band regions \cite{landau-lifschitz m} of the parameters of
\( \:
(\,|\vec{k}| \,, g\xi_{0}\,) \,,
\: \) 
or $(h\,, \theta )$.
It has been realized that this instability gives rise to particle production
\cite{baryogenesis review96}.
The most important recent development is that for large amplitude
oscillation particle production and associated field decay is
greatly expedited, typically ending in a time scale of 
O[10 --- 100] times $1/m_{\xi }$
\cite{reheating parametric}, 
\cite{linde et al 94}, \cite{holman 95}, \cite{mine95-96}, \cite{fkyy95}.
However, in all initial investigations so far no systematic estimate of the
environmental effect has been attempted.
This is precisely what we wish to do in the present section.
Throughout this section we assume that the time scale of cosmic
expansion is much larger than the oscillation time scale so that
the cosmological expansion may be ignored.
This is usually valid at the explosive stage of particle production.

This dynamical problem belongs to the more general class of dynamics of
time dependent harmonic oscillator.
We shall briefly mention how much of the general formalism
for the simple harmonic oscillator is modified in this case.
There are two time dependences in this system:
the nonlocal effect of the correlation $\alpha (\tau - s)$ introduced
by the environmental interaction, and the time dependent 
frequency $\omega (\tau )$.
The most important cases of applications can be analyzed using the approximate
localized friction for $\alpha _{I}(\tau - s)$, since we are primarily
interested in the late time behavior.
In this case one can explicitly perform the path integral in terms of
a classical solution for $q_{k}(\tau )$.

The notion of the localized friction arises by focusing on the late time
behavior of $\alpha _{I}(\tau )$. By the late time here
we mean the asymptotic
time that is much larger than any time scale of excitation in the
environment; $t \gg 1/$ (maximal excitation frequency) $\approx 1/\Omega $. 
The limiting form is then given by
\begin{eqnarray}
&&
\alpha _{I}(\tau )  = -\,\int_{0}^{\infty }\,d\omega 
\,r(\omega )\,\sin (\omega \tau ) \:\rightarrow  \: 
\eta \,\delta '(\tau ) + \delta \omega ^{2}\,\delta (\tau ) \,, \\
&&
\eta = -\,2\int_{0}^{\infty }\,d\tau \,\tau\,\alpha _{I}(\tau )
= \pi \,r'(0) \,, \hspace{0.5cm} {\rm for}\; 
r(\omega = \infty ) = 0 \,, \\
&&
\alpha _{R}(\tau ) \:\rightarrow  \: \frac{\eta }{\pi }\,
\int_{0}^{\infty }\,
d\omega \,\omega \,f(\frac{\omega }{\Omega })\,
\coth (\frac{\beta \omega }{2})\,\cos (\omega \tau )
\,.
\end{eqnarray}
The essence of this approximation is the low frequency truncation to
the response weight,
\begin{equation}
r(\omega ) = \omega \,r'(0) = \frac{\eta }{\pi }\,\omega \,.
\end{equation}

Although the dissipation given by $\alpha _{I}(\tau )$ is local by the
time scale larger than $1/\Omega $ the cutoff scale, 
the damping time scale of the noise kernel 
$\alpha _{R}(\tau )$ may differ. It can be shown that at high temperatures
of $T\geq \Omega $ the locality of the correlation holds for
$\tau \geq 1/\Omega $, giving
\begin{equation}
\alpha _{R}(\tau ) \sim \frac{\eta T}{\pi }\,\delta (\tau ) \,.
\end{equation}
On the other hand, at low temperatures of $T \leq \Omega $
the local approximation may be inadequate and a high frequency cutoff is
needed. 
As already discussed, we replace in this case 
the response weight 
by multiplying some cutoff function $f(x)$, 
for instance the simple frequency cutoff 
at $\omega = \Omega $ with $f(x) = \theta (1 - x)$, as is often practiced.

With $\alpha _{I}(\tau ) \:\propto  \: \delta '(\tau )$, the exponent
factor in the path integral becomes, including the mass renormalization
effect,
\begin{eqnarray}
&&
-\,
\frac{i}{2}\,\int_{0}^{t}\,d\tau \,
X(\tau )\,(\,\frac{d^{2}}{d\tau ^{2}}+ \omega _{R}^{2}
(\tau )\,)\,\xi(\tau )  + \frac{i}{2}\,\eta \,\int_{0}^{t}\,d\tau \,
X(\tau )\,\dot{\xi }(\tau ) 
\nonumber \\
&& 
+ \frac{i}{2}\,\left[\,X\dot{\xi }\,\right]^{t}_{0}
- \frac{i}{2}\,\eta \,\xi _{f}X_{f} \,{\rm sign}(t) 
- \int_{0}^{t }\,d\tau \,\int_{0}^{\tau }\,ds\,
\xi (\tau )\alpha_{R}(\tau - s)\xi (s) \,.
\end{eqnarray}
A remarkable feature of this formula is that  a part of the action 
excluding the $\alpha _{R}$ term is 
a local integral described in terms of a renormalized frequency 
$\omega _{R}(\tau )$ and the friction term $\propto \eta $.
This much is enough to considerably simplify calculation of the transition
amplitude of the $q-$system.

Path integral over the system variable $q$ is standard:
as in the more general case the equation for the classical path of
the difference path $\xi (\tau )$, 
which is obtained by functional differentiation
with respect to $X(\tau )$, is
\begin{eqnarray}
\left( \,\frac{d^{2}}{d\tau ^{2}} + \omega _{R}^{2}(\tau ) - 
\eta \,\frac{d}{d\tau }\,\right)\,\xi (\tau ) = 0 \,.
\end{eqnarray}
This equation is best analyzed by introducing a new $y(\tau )$,
\begin{eqnarray}
&&
y(\tau ) \equiv e^{\eta \tau /2}\,\xi (\tau ) \,, \\
&&
\left( \,\frac{d^{2}}{d\tau ^{2}} + \omega _{R}^{2}(\tau ) - 
\frac{\eta ^{2}}{4}\,\right)\,y(\tau ) = 0 \,.
\label{modified mathieu} 
\end{eqnarray}

The technique of the Laplace transform, which was very useful for the
simple harmonic oscillator in the previous sectios, 
does not work here due to the time dependence $\omega_{R}^{2}(\tau )$.
This is another main reason we have to resort to the local friction
approximation.

Limiting our analysis to modes within the instability band of the modified
Mathieu equation, we note that
two linearly independent solutions are either growing or decaying
according to
\begin{equation}
y(\tau ) = e^{\lambda m_{\xi }\tau }\,P(\tau ) \,, \hspace{0.5cm} 
{\rm or} \hspace{0.5cm} 
e^{-\,\lambda m_{\xi }\tau }\,R(\tau ) \,, 
\end{equation}
where $\lambda > 0$ is the dimensionless growth rate and both $P(\tau )$ and
$R(\tau )$ are periodic with period of $ 2\pi /m_{\xi }$.
There are two independent solutions for $y(t)$,
\( \:
u(t) 
\: \) and $v(t)$
satisfying the boundary condition,
\( \:
u(0) = 0 \,, \; v(t) = 0 \,.
\: \)
These are linear combinations of the growing and the decaying
solutions. With the normalization appropriately given,
leading asymptotic behaviors of these are
\begin{equation}
u(t) \sim e^{\lambda m_{\xi }t}\,\tilde{P}(t) \,, \hspace{0.5cm} 
\dot{v}(t) = - \,\frac{\dot{u}(0)v(0)}{u(t)} \sim 
e^{-\,\lambda m_{\xi }t}\,\tilde{R}(t) \,,
\end{equation}
with both $\tilde{P}(t)$ and $\tilde{R}(t)$ bounded functions.

Note that the growth rate $\lambda $ here is related to the solution
of the modified Mathieu equation, eq.(\ref{modified mathieu}), and
not of the original one.
Thus we should keep in mind that the growth rate $\lambda $ does
depend on the friction $\eta $:
\( \:
\lambda = \lambda (\eta ) \,.
\: \)
Suppose for instance that the renormalized $\varphi-$mass vanishes:
\( \:
m_{\varphi }^{2} + \delta \omega ^{2} = 0 \,.
\: \)
The $\eta $ term in eq.(\ref{modified mathieu}) then has an effect of
lowering the band level, since the $h$ parameter in the standard Mathieu
equation is modified to
\begin{equation}
\tilde{h} = h - \frac{\eta ^{2}}{4m_{\xi }^{2}}
= \frac{\vec{k}^{2}}{m_{\xi }^{2}} + 2\theta 
- \frac{\eta ^{2}}{4m_{\xi }^{2}} \,, 
\end{equation}
in this case.
Although it is in general difficult to estimate the influence of 
the friction $\eta $
on the growth rate $\lambda (\eta )$, the general trend is obvious:
the lowering effect of the band level tends to increase the growth rate.
In the extreme case of a very large $\eta $ the $h$ parameter becomes
negative, and the parameter is clearly in the instability region.
In perturbation theory we later work with, the leading
$\lambda $ term is however independent of $\eta $.

The exponential growth of the original mode function $q_{k}(\tau )$
is possible for \\
$2\lambda(\eta )\, m_{\xi } > \eta $.
We observe two competitive factors for the growth:
the rate of the parametric amplification $2\lambda(\eta )\, m_{\xi }$ 
against the friction $\eta $.
As will be shown in subsequent computations of physical quantities, 
the friction, when it is small, does
act to diminish the parametric particle production, but does not
wipe out the parametric effect. Thus, instead of the blocking factor,
it is more appropriate to view  the friction as
the inverse time scale for the system to be driven towards thermalization.
Hence if the friction is small enough, the parametric amplification
never loses against the friction.

In subsequent discussion we only present analytic computation
of time evolution of physical quantities and leave results of numerical 
calculation to our paper \cite{hjmy 96}. 
In analytic estimate we concentrate on the asymptotic late time behavior 
for which
\( \:
\tau \gg {\rm max.} \;(\,1/\eta \,, 1/(\lambda m_{\xi })\,)
\: \).
With the asymptotic behavior of the classical solution, $u(\tau ) $ and $ 
v(\tau )$, the three quantities that appear in the reduced density matrix 
(\ref{density matrix}) of the $\varphi-$system are
\begin{eqnarray}
{\cal A} &\rightarrow & \frac{1}{8K}\,(\frac{\dot{u}(0)}{u(t)})^{2}
\,, 
\\ 
{\cal B} &\rightarrow&  \frac{A}{2} - \frac{C^{2}}{2K} \,, 
\\ 
{\cal C} &\rightarrow  &
\frac{C}{2K}\, \frac{\dot{u}(0)}{u(t)}
- \frac{1}{2}\, \left( \,\frac{\eta }{2} - \frac{\dot{u}(t)}{u(t)}\,\right)
\,.
\end{eqnarray}
The quantities $K\,, A \,, C$ are defined by
\begin{eqnarray}
K &=& B + e^{-\,\eta t}\,
\frac{\omega _{0}}{2}\,\coth  (\frac{\beta _{0}\omega _{0}}{2})
\left( \, 1 + \frac{1}{\omega _{0}^{2}}\,
(\,\frac{\eta }{2} + \frac{\dot{v}(0)}{v(0)}\,)^{2}
\,\right) \,, \label{k-equation} \\
(\,A \,, B \,, C\,) 
&=& \int_{0}^{\infty }\,d\omega \,\coth (\frac{\beta \omega }{2})\,
r(\omega )\,(a\,, b\,, c)(\omega )  \,, \\
a(\omega ) &=&
\left|\,\int_{0}^{t}\,ds\,\frac{u(s)}{u(t)}\,
e^{i\omega s - \frac{\eta }{2}(t - s)}\,\right|^{2} \,, 
\\ 
b(\omega ) &=& 
\left|\,\int_{0}^{t}\,ds\,\frac{v(s)}{v(0)}\,
e^{i\omega s - \frac{\eta }{2}(t - s)}\,\right|^{2} 
\,, \\ 
c(\omega ) &=& \Re \,\left( \,
\int_{0}^{t}\,ds\,\frac{u(s)}{u(t)}\,
e^{i\omega s - \frac{\eta }{2}(t - s)}\cdot 
\int_{0}^{t}\,ds\,\frac{v(s)}{v(0)}\,
e^{-i\omega s - \frac{\eta }{2}(t - s)}\,\right) \,.
\label{last c-eq}
\end{eqnarray}
They have asymptotic behaviors:
\begin{equation}
A = O[1] \,, \hspace{0.5cm} C \leq  
O[\,{\rm Max.\;} (\,e^{-\,\lambda m_{\xi }t}\;,e^{-\,\eta t/2})\,] 
\,, \hspace{0.5cm} 
K = O[\,{\rm Max.\;} (\,e^{-\,2\lambda m_{\xi }t}\;,e^{-\,\eta t})\,] \,.
\end{equation}
Thus, we conclude that ${\cal A} \,, {\cal B} \,, 
{\cal C}$ behave asymptotically as
\begin{equation}
{\cal A} = O[\,{\rm Min.}\;(\,1\,, e^{-\,2\lambda m_{\xi }t + \eta t}\,)\,] 
\,, \hspace{0.5cm} {\cal B} = O[1] \,, \hspace{0.5cm} 
{\cal C} = O[1] \,.
\end{equation}
We omitted dimensional scales of these quantities such as
$\omega _{0}$ and $\eta $, since these differ from case to case,
depending on the parameter range of
\( \:
T \,, \; \omega _{0} \,, \; m_{\xi } \,, \; \eta \,.
\: \)
For the reference frequency we take
\( \:
\omega _{0}^{2} = \omega _{R}^{2}(0) \,:
\: \)
the initial value of the frequency.

The formulas so far are valid for any response weight $r(\omega )$.
We now specialize to the environment system that may be described by
the approximate form of the localized friction,
\( \:
r(\omega ) = \frac{\eta }{\pi }\,\omega \; \times 
\: \)
(high $\omega-$cutoff).
In particular, we are very much interested in the 
high and low temperature limits of various quantities.
The high temperature here effectively means that 
\( \:
T \gg {\rm Max.}\;(\,\omega _{0} \,, m_{\xi } \,, \eta \,, \Omega \,)
\,.
\: \)

Here we shall write only some typical results for the average energy,
the entropy, and the distribution function:
for asymptotic late times we find at high temperatures
\begin{eqnarray}
\langle \omega  \rangle &\approx  & \omega _{0}\, \langle n \rangle 
\; \approx \;
\frac{\eta \omega _{0}^{2}\,T}{2\lambda m_{\xi }}\,
e^{-\,\eta t}\,(\frac{u(t)}{\dot{u}(0)})^{2}\,
\left( \,1 + 
\frac{1}{\omega _{0}^{2}}\,(\,\frac{\eta }{2} - \frac{\dot{u}(t)}{u(t)}\,)^{2}
\,\right) 
\,, \\
S &\approx &
\ln \,\left( \,\frac{\eta T}{\lambda m_{\xi }}\,|\frac{u(t)}{\dot{u}(0)}|
\,e^{-\eta t/2}\,\right)
\,, \\
f(\omega ) &\approx &
\frac{1}{\sqrt{\,2\pi \,\langle \omega  \rangle\,}}\,
\frac{e^{-\,\omega /(\,2\langle \omega  \rangle\,)}}{\sqrt{\omega }} \,.
\label{asymptotic dist}
\end{eqnarray}
Both the number $\langle n \rangle$ and the average energy
$\langle \omega  \rangle$ increase exponentially with time, the rate
$2\lambda m_{\xi } - \eta $ 
being somewhat diminished from the value taken without environment.
Note also that a rather non-trivial enhancement factor $\:\propto  \: T$
appears in the prefactor.

Despite of the suppression
of the growth rate from $2\lambda m_{\xi }$ to $2\lambda m_{\xi } - \eta $,
the environment interaction does not erase the parametric effect if
\begin{equation}
2\lambda(\eta )\, m_{\xi } > \eta \,.
\label{parametric condition} 
\end{equation}
Let us look into the meaning of this condition
that guarantees the exponential rate of particle production
in thermal bath.
Since $\eta $ is the relaxation rate of $\varphi-$field disturbance
towards thermalization,
the condition simply implies that the thermal relaxation 
never catches up the parametric amplification for a sufficiently
small friction.
At the same time the condition implies
that there exists a critical strength of the friction $\eta $
above which the parametric effect does not occur.
The equation 
\( \:
2\lambda (\eta )\,m_{\xi } = \eta 
\: \)
thus gives the critical strength of the friction constant $\eta _{c}$.
We wish to stress that although the suppression of the growth rate might
have been anticipated from a naive consideration \cite{reheating parametric},
it was not clear in the past 
how one can verify the parametric amplification
itself, starting from quantum mechanical principles.
In our view the significance of the friction $\eta $ is in its
role as the relaxation rather than destruction of the coherence.
Hence if the relaxation time scale is larger than the amplification time
scale, the parametric effect wins over the decoherence due to the environment.

We carried out detailed numerical analysis in \cite{hjmy 96}.
It supports analytical estimates.
We only quote our main conclusions from that work:
\begin{itemize}
\item 
For the first time it was verified that the parametric amplification
can occur in medium, using the influence functional method.
The average energy and the total number of population
exponentially increases with time, if the growth rate of the parametric
amplification in medium is larger than the friction $\eta $,
interpreted to be a measure of the relaxation rate towards thermalization
of the system variable.
The critical friction for the exponential growth has also been given, 
above which the parametric effect does not occur.
\item 
The resulting energy distribution deviates from the thermal Boltzmann
distribution, having a long tail of the high energy component 
characterized by an exponentially increasing average energy and
an exponentially decreasing $\delta $ parameter(spectrum shape parameter).
\item 
Late time behavior of the average energy, the entropy, and the
energy spectrum is insensitive to the environment temperature
if time is displaced in comparison at different temperatures. 
This is true, even including the $T = 0 $ vacuum.
\end{itemize}

\vspace{0.5cm} 
\begin{center}
{\bf Acknowledgement}
\end{center}

The results presented in this report have grown from works in
collaboration with M. Hotta, I. Joichi, and Sh. Matsumoto.
I wish to thank all of them for extensive discussions.
This work has been supported in part by the Grand-in-Aid for Science
Research from the Ministry of Education, Science and Culture of Japan,
No. 08640341 and No. 07304033.

\end{document}